\documentclass[twocolumn]{aastex631}

\usepackage{color}

\newcommand{\kms}{\ifmmode {\rm km\ s}^{-1} \else km s$^{-1}$\fi}
\newcommand{\ergs}{\ifmmode {\rm erg\ s}^{-1} \else erg s$^{-1}$\fi}
\newcommand{\lb}{\ifmmode L_{\rm Bol} \else $L_{\rm Bol}$\ \fi}
\newcommand{\ledd}{\ifmmode L_{\rm Edd} \else $L_{\rm Edd}$\ \fi}
\newcommand{\lx}{\ifmmode L_{\rm 2-10keV} \else  $L_{\rm 2-10keV}$\ \fi}
\newcommand{\ha}{\hbox{H$\alpha$}}
\newcommand{\hb}{\hbox{H$\beta$}}

\newcommand{\mbh}{\ifmmode M_{\rm BH}  \else $M_{\rm BH}$\ \fi}
\newcommand{\lv}{\ifmmode \lambda L_{\lambda}(5100\Ang) \else $\lambda L_{\lambda}(5100\Ang)$\ \fi}
\newcommand{\lbol}{\ifmmode L_{\rm Bol} \else $L_{\rm Bol}$\ \fi}

\newcommand{\oi}{\hbox{[O\,{\sc i}]}}
\newcommand{\oii}{\hbox{[O\,{\sc ii}]}}
\newcommand{\nii}{\hbox{[N\,{\sc ii}]}}
\newcommand{\sii}{\hbox{[S\,{\sc ii}]}}
\newcommand{\oiii}{\hbox{[O\,{\sc iii}]}}

        \newcommand{\msun}{ M_{\odot}}

\newcommand{\hii}{\hbox {H\,{\sc II}}}

\newcommand{\oh}{\ifmmode 12+ \log({\rm O/H}) \else 12+log(O/H) \fi}
\newcommand{\mdot}{\ifmmode \dot{m} \else \dot{m} \fi }
\newcommand{\llog}{\ifmmode {\rm log} \else {\rm log} \fi }
\newcommand{\Ang}{\mathring{\mathrm{A}}}

\begin{document}

\title{Clumpy Starburst in a Local Dwarf Galaxy, NGC 1522}

\correspondingauthor{Qiusheng Gu, Yulong Gao}
\email{qsgu@nju.edu.cn, yulong@nju.edu.cn}

\author[0009-0003-6303-7329]{Liuze Long}
\affiliation{School of Astronomy and Space Science, Nanjing University, Nanjing 210093, China}
\affiliation{Key Laboratory of Modern Astronomy and Astrophysics (Nanjing University), Ministry of Education, Nanjing 210093, China}

\author{Yulong Gao}
\affiliation{School of Astronomy and Space Science, Nanjing University, Nanjing 210093, China}
\affiliation{Key Laboratory of Modern Astronomy and Astrophysics (Nanjing University), Ministry of Education, Nanjing 210093, China}

\author{Qiusheng Gu}
\affiliation{School of Astronomy and Space Science, Nanjing University, Nanjing 210093, China}
\affiliation{Key Laboratory of Modern Astronomy and Astrophysics (Nanjing University), Ministry of Education, Nanjing 210093, China}

\author{Yong Shi}
\affiliation{School of Astronomy and Space Science, Nanjing University, Nanjing 210093, China}
\affiliation{Key Laboratory of Modern Astronomy and Astrophysics (Nanjing University), Ministry of Education, Nanjing 210093, China}

\author{Xin Li}
\affiliation{School of Astronomy and Space Science, Nanjing University, Nanjing 210093, China}
\affiliation{Key Laboratory of Modern Astronomy and Astrophysics (Nanjing University), Ministry of Education, Nanjing 210093, China}

\author{Can Xu}
\affiliation{School of Astronomy and Space Science, Nanjing University, Nanjing 210093, China}
\affiliation{Key Laboratory of Modern Astronomy and Astrophysics (Nanjing University), Ministry of Education, Nanjing 210093, China}

\author{Yifei Jin}
\affiliation{Harvard and Smithsonian Center for Astrophysics, 60 Garden Street, MA 02138}

\author{Zhiyuan Zheng}
\affiliation{School of Astronomy and Space Science, Nanjing University, Nanjing 210093, China}
\affiliation{Key Laboratory of Modern Astronomy and Astrophysics (Nanjing University), Ministry of Education, Nanjing 210093, China}

\author{Jing Dou}
\affiliation{School of Astronomy and Space Science, Nanjing University, Nanjing 210093, China}
\affiliation{Key Laboratory of Modern Astronomy and Astrophysics (Nanjing University), Ministry of Education, Nanjing 210093, China}

\author{Fuyan Bian}
\affiliation{European Southern Observatory, Alonso de C{\'o}rdova 3107, Casilla 19001, Vitacura, Santiago 19, Chile}

\author{Xiaoling Yu}
\affiliation{College of Physics and Electronic Engineering, Qujing Normal University, Qujing 655011, Yunnan Province, China}

\begin{abstract}

To investigate the star-forming process in nearby dwarf galaxies, we present Integral Field Units (IFU) observation of the star-forming dwarf galaxy, NGC 1522, with the Very Large Telescope (VLT)/Multi Unit Spectroscopic Explorer (MUSE) as a part of Dwarf Galaxy Integral Survey (DGIS). Our observation reveals the presence of a star-forming clumpy ring in its central region. We identify nine distinct star-forming clumps based on extinction-corrected $\ha$ emission-line map, with the total star formation rate (SFR) of about 0.1 $M_\odot$ yr$^{-1}$. The nine clumps are considered to be starbursts, which represent an extreme case in the local universe, without invoking major merging. We investigate the properties of ionized gas using the strong emission lines and `BPT' diagrams, in conjunction with the velocity mapping. Our analysis unveils intriguing patterns, including the positive metallicity gradient and low N/O abundance ratio. This peculiar distribution of metallicity may signify external gas accretion. Our results suggest that the ongoing star formation in NGC 1522 might be triggered and sustained by the inflow of external metal-poor gas.

\end{abstract}


\section{Introduction} \label{sec:intro}

Dwarf galaxies, characterized by their low mass and wide distribution, play a pivotal role in galaxy formation and evolution. In the framework of the $\Lambda$ Cold Dark Matter ($\Lambda$CDM) cosmology, these small systems gradually merge to form larger ones \citep{1994MNRAS.267..981K,2006Natur.440.1137S}. Consequently, dwarf galaxies are often considered to be building blocks of massive galaxies, prompting extensive studies of their properties. Among these diminutive stellar systems, blue compact dwarf galaxies (BCDs) stand out as especially low-mass, compact entities with exceptionally high star formation rates \citep[SFR,][]{2003ApJS..147...29G,2016A&A...585A..79F}. They are also characterized by their gas-rich nature \citep{2002AJ....124..191S} and relatively low metallicity \citep{2000A&ARv..10....1K}, making them ideal targets for unraveling the intricate processes of star formation within dwarf galaxies.

Unlike massive galaxies with regular structures such as spiral arms, BCDs lack significant pattern features, leaving open questions about the mechanisms that initiate and sustain their star formation. In low-mass systems, the conversion of baryons into stars is hindered by weak gravitational potentials \citep{2007MNRAS.382.1187K}. Alternative mechanisms, such as gravitational interactions, mergers, or gas inflow/outflow, are proposed to enhance star formation  \citep{2014ApJ...789L..16L,2015ApJ...807L..16K,gao2022a,gao2023}. Given the hierarchical nature of structure formation, interactions among dwarf galaxies are common in the early Universe \citep{2018MNRAS.480.3376B}, however, in the local universe, most dwarf galaxies either exist as satellites or isolated systems. It is proposed that the accretion of cold gas may be a significant driver of star formation in isolated BCDs \citep{2023MNRAS.522.2089D}.

Nearby BCDs are exceptional candidates for detailed investigations of processes of star formation, as they offer an opportunity to study both their gas and physical young stellar components \citep[e.g.,][]{2020A&A...635A.134D}. The unique combination of low metal content, high SFR, and clumpy morphology \citep{2003ApJS..147...29G} aligns BCDs with the characteristics of high-redshift galaxies \citep[e.g., ][for a review]{2022NatAs...6...48A}. The prevalence of off-center starburst clumps, indicative of intense star formation episodes triggered by gas accretion, has been observed in intermediate and high-redshift galaxies \citep{2011Msngr.145...39F,Genel_2012,2012ApJ...752..111N,2023arXiv230800041M}. Nearby BCDs also exhibit signs of star formation in their outer regions \citep{2015ApJ...810L..15S}, while kinematic analysis suggests that erratic velocities may stem from mergers and/or infall of metal-poor gas, driving the vigorous star formation \citep{2016MNRAS.462.3314W}.

Moreover, BCDs undergo a dynamic equilibrium between gas accretion, outflows, and intense star formation. However, direct observations of gas accretion and outflows remain challenging \citep{2014A&ARv..22...71S}. \citet{2018ApJ...867..142C} investigated metal accumulation in dwarf galaxies, through tracing the history of metal enrichment and outflows.  Metallicity is an integrated physical parameter in galaxy evolution studies because it reflects the processes governing the baryon cycle within galaxies. The association between low metallicity and BCDs suggests that these systems represent some of the earliest and most youthful objects in the universe. By scrutinizing BCDs, we aim to uncover insights into the formation and evolution of dwarf galaxies, shedding light on the characteristics of the first galaxies that emerged during the universe's infancy.

In the past decade, Integral Field Units (IFU) spectrographs have revolutionized our ability to conduct 3D spectroscopic analyses, providing unprecedented spatial resolution, and surpassing the capabilities of traditional long-slit observations. Using the data from the Mapping Nearby Galaxies at Apache Point Observatory (MaNGA) survey, \citet{2019ApJ...872..144H} observed both the O/H and N/O abundance ratios, unveiling anomalously low metallicity in select star-forming regions. Additionally, \citet{2016A&A...585A..79F} performed observations of a colliding dwarf galaxy with the Multi Unit Spectroscopic Explorer (MUSE) at the Very Large Telescope (VLT), revealing regions with atypical $\oi$ emission lines encircling the galaxy. \citet{2023MNRAS.522.2089D} conducted a chemodynamical study of a nearby, gas-rich starburst dwarf, identifying an anti-correlation between SFR and gas metallicity, suggesting recent off-center starbursts may originate from the accretion of metal-poor gas. VLT/MUSE, in particular, with its exceptional combination of high spatial and spectral resolution, a broad field of view, and an extended wavelength range, has emerged as a powerful instrument for investigating BCDs.

This study will focus on the nearby BCD, NGC 1522 \citep{2003ApJS..147...29G,2005ApJS..156..345G}. NGC 1522 has garnered attention as an enigmatic lenticular (S0) galaxy \citep{2018MNRAS.480.4931V}. The best-fit model derived from the \texttt{STARLIGHT} analysis indicates that 95$\%$ of its flux can be attributed to a stellar population younger than 0.1 Gyr. Notably, NGC 1522's long-slit spectroscopy reveals an array of strong emission lines, including $\ha$, indicative of $\hii$ regions within the galaxy.

In this study, we present the VLT/MUSE \citep{Bacon_2010} IFU observations to investigate the properties of ionized gas in the dwarf galaxy NGC 1522. 
The paper is structured as follows: In Section~\ref{sec:data}, we introduce the observations and describe the data reduction process, followed by a presentation of the star formation properties. Section~\ref{result} offers a comprehensive analysis of emission lines, with a particular emphasis on spatial variations. The ensuing discussions are presented in Section~\ref{discussion}, and we draw our final conclusions in Section~\ref{conclusion}. Throughout this paper, we adopt a flat $\Lambda$CDM cosmology model with the following parameters: $\Omega_\Lambda=0.7, \Omega_M=0.3, $ and $H_0 = 70$ km $\cdot $ s$^{-1} \cdot$ Mpc$^{-1}$.

\section{Data}
\label{sec:data}
In this section, we present our observation and data reduction of optical spectroscopy. We aim to infer the ionized and chemical properties of NGC 1522. The archival parameters of NGC 1522 are summarized in Table~\ref{1522Prop}, among of which the inclination is obtained by fitting IRAC/Spitzer 3.6$\mu$m image using \texttt{photutils} package\footnote{\url{https://photutils.readthedocs.io}}.

\subsection{VLT/MUSE data}

NGC 1522 was observed using VLT/MUSE in November 2019 to cover the central $63''\times 63''$ regions, as a part of the Dwarf Galaxy Integral Survey (DGIS, \citet{2025arXiv250104943L}; PI: Yong Shi and Fuyan Bian). The total exposure time is approximately 1.2 hours. First, the raw data is automatically reduced by MUSE Data Reduction Software \footnote{\url{https://www.eso.org/sci/software/pipelines/muse/}}.  To improve the flux calibration, we remove the residual sky for individual frame as post-processing. Finally, the datacube was rebinned by 2$\times$2 pixels, and World Coordinate System (WCS) are corrected by comparing the brightest point with DESI image.
The observations exhibited a seeing value of approximately 0.5\arcsec. The wavelength range spans from 4600 \AA\ to 9350 \AA, with a spectral resolution of 1.25 \AA.

To analyze the stellar continuum, we employed the Penalized PiXel-Fitting package  \citep[ \texttt{pPXF},][]{2004PASP..116..138C,2017MNRAS.466..798C,Cappellari2022}, a Python implementation method. We utilized the MILES simple stellar population (SSPs) templates \citep{2006MNRAS.371..703S,2010MNRAS.404.1639V}, assuming \citet{1955ApJ...121..161S} initial mass function (IMF) and \citet{2000ApJ...533..682C} dust extinction curve. All spectra were fitted within the wavelength range of 4600 \AA\ to 9000 \AA. To derive pure emission line fluxes, we subtracted the stellar contribution from the processed spectrum and employed a single Gaussian fit for each strong emission line, including $\ha$, $\hb$ $\oiii\lambda\lambda4959, 5007$, etc. To ensure the reliability of our results, we considered only spatial pixels (spaxels) with Signal-to-Noise Ratios (S/N) larger than 5 for $\ha$, $\hb$ and $\oiii\lambda\lambda4959, 5007$, and S/N $\geq$ 3 for $\oi\lambda6300$, $\nii\lambda6583$ and $\sii\lambda\lambda6716,6731$. 

\begin{figure*}
	\centering
	\includegraphics[width=1\columnwidth]{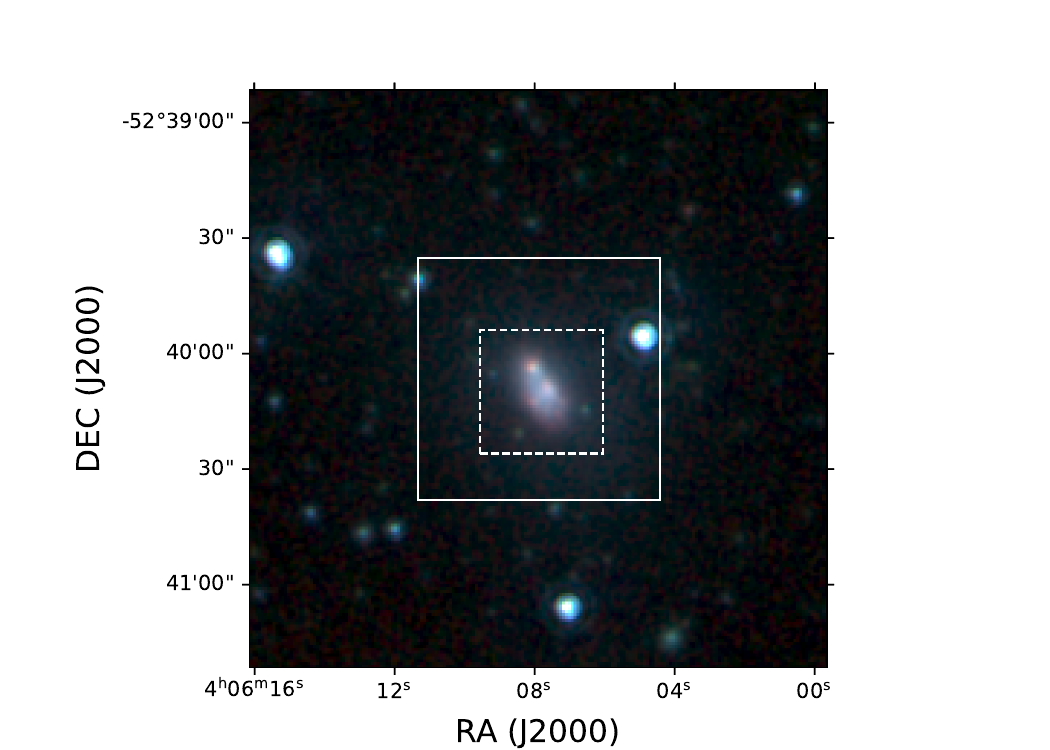}
	\includegraphics[width=0.98\columnwidth]{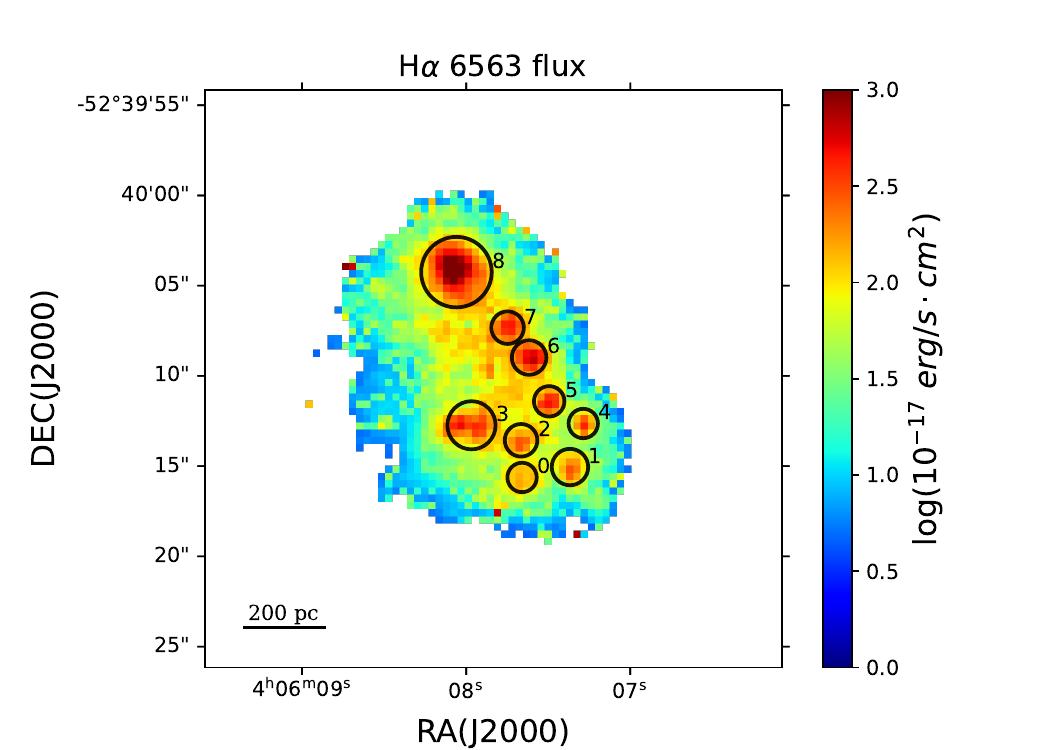}
	\caption{\textit{Left:} The pseudo-color image (combined with the IRAC/Spitzer 3.6$\mu$m, 4.5$\mu$m and 5.8$\mu$m images) of NGC 1522. The white square (dash line) shows the analysis region we cut out from the field-of-view (FoV) region (white solid line) observed by MUSE. \textit{Right:} The integrated intensity map of the extinction-corrected $\ha$ emission line (subtracted the stellar contribution). The black circles (solid line) indicate the 9 star-forming clumps identified using \texttt{Astrodendro}. Around the circles, we also label the IDs of these clumps.}
	\label{knots}
\end{figure*}

\begin{table}
    \centering
    \caption{Properties of NGC 1522} \label{1522Prop}  
    \begin{tabular}{lrr}
        \hline
        Property &  \\
        \hline
        RA (J2000) & 04h06m07.92s     \\
        DEC (J2000)& -52d40m06.30s \\
        Distance/Mpc$^{(1)}$ & 9.3\\
        Inclination angle/$^\circ $ & 59.6 \\
        $z^{(2)}$   &   0.0030  \\
        12 + $\log$(O/H)$^{(3)}$ & 8.20 $\pm$ 0.12 \\
        $\log_{10}(M_*/\msun)^{(4)}$ & 8.32 $\pm$ 0.05\\
        SFR/$\msun \rm \ yr^{-1}$ & 0.098 $\pm$ 0.003 \\
        $\log_{10}(M_{HI}/\msun)^{(5)}$&8.40\\
        \hline
    \end{tabular}
    \tablecomments{(1) \citet{Lee_2011}; (2) \citet{2014ApJ...783..135A}; (3) \citet{1994ApJ...420..576M}; (4) \citet{2023AJ....165..260D}; (5) \citet{2011AJ....142...83N}.}
\end{table}

\subsection{Star-forming clumps}
\label{SFknots}

With the high spatial resolution 2-D spectra provided by MUSE, we find that star formation is mainly concentrated at individual clumps in the central region. We employed the \texttt{Astrodendro}\footnote{\url{https://dendrograms.readthedocs.io}} Python package \citep{2009Natur.457...63G} to detect these star-forming clumps in the $\ha$ emission line map. This dendrogram-based algorithm is well-suited for identifying robust $\ha$-bright regions in galaxies \citep[e.g.,][]{2020A&A...635A.134D}. The \texttt{Astrodendro} algorithm defines the boundaries of clump structures using specific parameters, including \textit{min$\_$value}, \textit{min$\_$delta}, and \textit{min$\_$npix}, where \textit{min$\_$value} represents the minimum flux value considered during the clump search, \textit{min$\_$delta} sets the minimum significance required to avoid including local maximum, and \textit{min$\_$npix} establishes the minimum number of pixels that a structure must contain. In our analysis, we adopted the smallest region radius based on the seeing value and set \textit{min$\_$npix} to 6. We finally identified nine distinct star-forming clumps in NGC 1522 located 50 to 120 pc away from the center, as illustrated in the right panel of Figure~\ref{knots}. These clumps show a ring-like structure with a deprojected radius of about 300 pc. Additionally, the left panel of Figure~\ref{knots} provides a pseudo-color image created by combining the IRAC/Spitzer 3.6$\mu$m, 4.5$\mu$m and 5.8$\mu$m images. The white square (solid line) represents the field-of-view (FoV) region coverd by MUSE, and the white square (dash line) highlights the analysis region of NGC 1522. We summarize our results of these clumps in Table~\ref{knotsprob} based on their observed optical spectra.

\begin{table*}
	\centering 
	\caption{Properties of star-forming clumps.}
	\label{knotsprob}
	\begin{tabular}{cccrcrrr}
		\hline
    	ID & RA & Dec & Radius& SFR &	12+log(O/H)$_{\text{N2}}$  & 12+log(O/H)$_{\text{O3N2}}$  &  log(N/O)  \\
    	&(J2000)&(J2000)&(pc)&($10^{-3}\msun$ yr$^{-1}$)& & & \\
		\hline	 
	    0 & 04h06m07.66s & -52d40m15.63s & 50.5 & 0.965 $\pm$ 0.001  &  8.267 $\pm$ 0.094 & 8.147 $\pm$ 0.081 & -1.395 $\pm$ 0.318  \\  
	    1 & 04h06m07.37s & -52d40m15.05s  & 62.6 &  1.460 $\pm$ 0.002 &  8.285 $\pm$ 0.094 & 8.134 $\pm$ 0.081 & -1.430 $\pm$ 0.317\\ 
	    2 & 04h06m07.66s & -52d40m13.57s & 57.0 & 1.393 $\pm$ 0.002 & 8.249 $\pm$    0.094 & 8.113 $\pm$ 0.081 & -1.425 $\pm$ 0.317\\      
	    3 & 04h06m07.97s & -52d40m12.74s & 82.8 &  3.917 $\pm$ 0.003  &  8.242 $\pm$ 0.092 & 8.124 $\pm$ 0.081 & -1.406 $\pm$ 0.314\\         
	    4 & 04h06m07.29s & -52d40m12.64s & 50.5 & 1.064 $\pm$ 0.002  &  8.174 $\pm$  0.095 & 8.084 $\pm$ 0.082 & -1.334 $\pm$ 0.321 \\    
	    5 & 04h06m07.49s & -52d40m11.41s & 52.4  & 1.832 $\pm$ 0.002  & 8.156 $\pm$   0.094 & 8.072 $\pm$ 0.081 & -1.376 $\pm$ 0.317 \\
	    6 & 04h06m07.62s &-52d40m08.98s & 59.4 & 2.940 $\pm$ 0.003  & 8.333 $\pm$  0.092 & 8.179 $\pm$ 0.081 & -1.432 $\pm$ 0.313\\    
	    7 & 04h06m07.75s & -52d40m07.33s & 57.0 & 2.271 $\pm$ 0.002  & 8.217 $\pm$    0.092 & 8.103 $\pm$ 0.081 & -1.361 $\pm$ 0.315  \\
	    8 & 04h06m08.06s &-52d40m04.25s & 121.2 & 17.052 $\pm$ 0.008 & 8.207 $\pm$ 0.091 & 8.103 $\pm$ 0.081 & -1.320 $\pm$ 0.312 \\    
	    \hline	 
	\end{tabular}
\end{table*}

\section{Results}\label{result}

With the advent of IFU observation, high-resolution images of ionized gas provide new insights into gas kinematics and star formation activity, enabling detailed spatial characterization within dwarf galaxies. In this section, we present the spatial distribution of ionized gas within NGC 1522 and derive the chemical properties of the star-forming clumps for the first time.

\subsection{Emission-line ratios and BPT diagrams}\label{subsec:bpt}

\begin{figure*}
	\includegraphics[width=1\columnwidth]{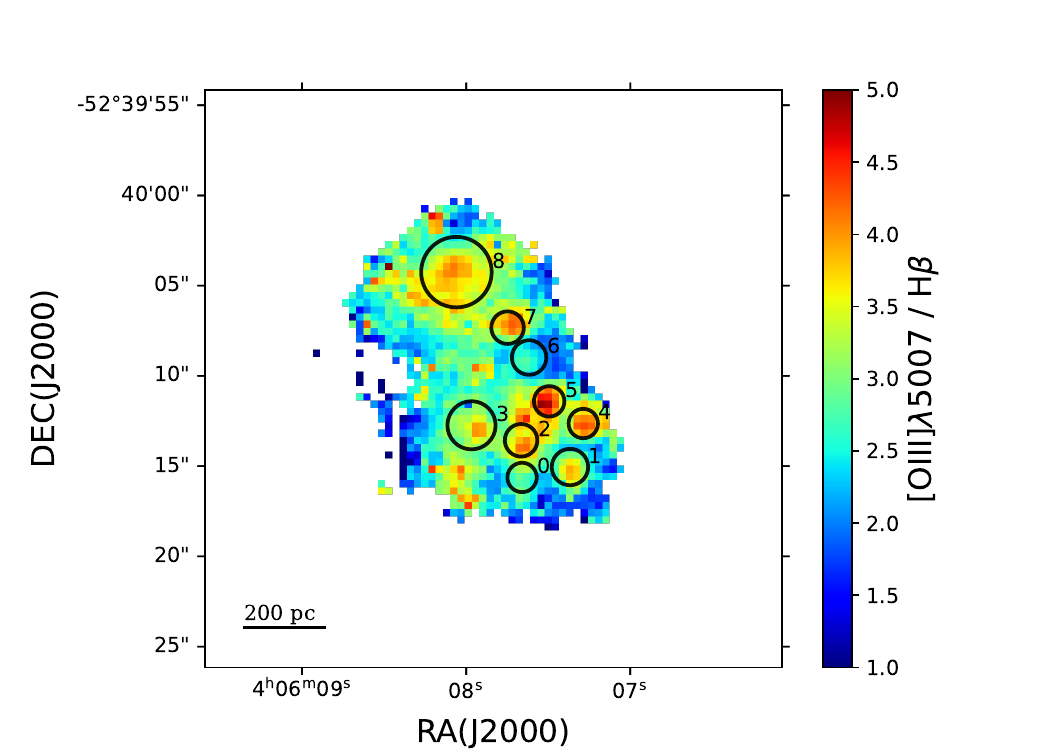}
    \includegraphics[width=1\columnwidth]{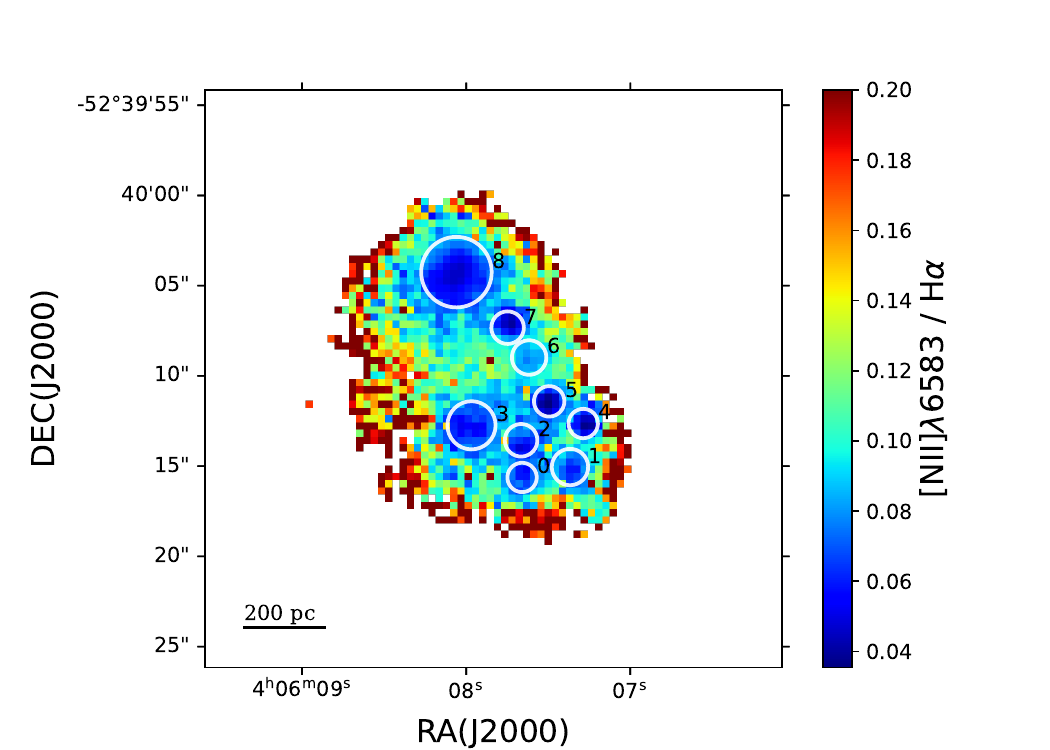}
    
    \includegraphics[width=1\columnwidth]{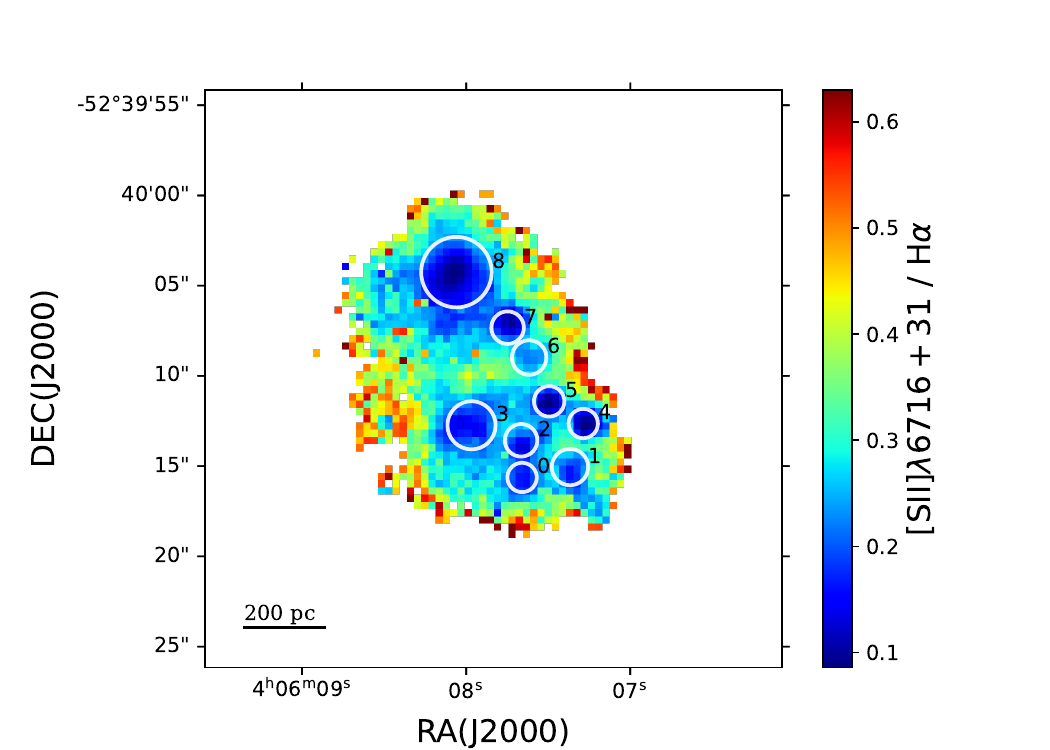}
    \includegraphics[width=1\columnwidth]{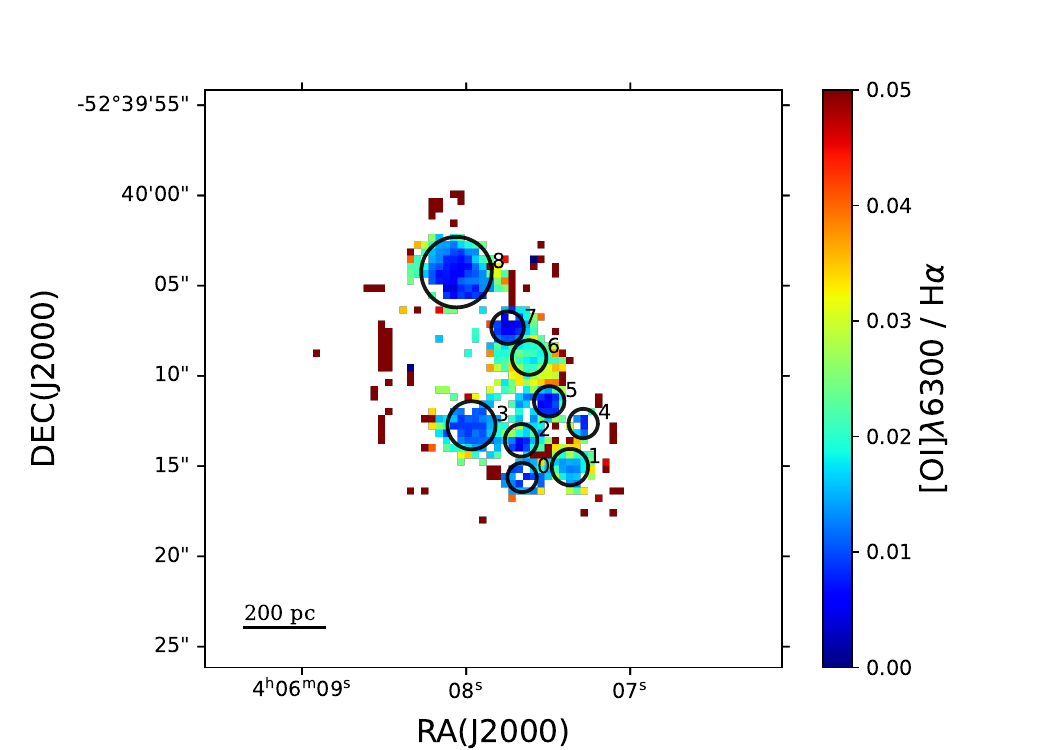}
    
    \caption{Spatial distribution of four primary emission line flux ratios: $\oiii$5007/$\hb$ (upper-left), $\nii$6583/$\ha$ (upper-right), $\sii$6716 + 31/$\ha$ (bottom-left), and $\oi$6300/$\ha$ (bottom-right). All emission lines are presented after correcting for attenuation. The black(white) circles mark the locations of the identified star-forming clumps. The field of view matches that of Figure~\ref{knots}.}
    \label{emission map}
\end{figure*}

Using the Calzetti dust extinction law \citep{2000ApJ...533..682C} and the Case B recombination model, we corrected the emission line fluxes for dust extinction, by using the Balmer decrement.
All subsequent emission line fluxes have been adjusted for dust attenuation.
In Figure~\ref{emission map}, we display maps of emission line ratios of ionized gas, including $\oiii\lambda 5007$/$\hb$, $\nii\lambda 6583$/$\ha$, $\sii\lambda 6716$/$\ha$, and $\oi\lambda6300$/$\ha$. Overlaid in blue circles are the star-forming clumps identified in Section~\ref{SFknots}. These maps provide insights into the ionization levels within NGC 1522's $\hii$ regions and enable the calculation of gas-phase metallicity, as discussed in Section~\ref{Sec_metal}. Notably, NGC 1522 exhibits an inside-out positive gradient in the $\nii\lambda 6583$/$\ha$, $\sii\lambda 6716$/$\ha$, and $\oi\lambda6300$/$\ha$ line ratio maps. The peaks in $\oiii$/$\hb$ coincide with regions displaying lower $\sii$/$\ha$ values, a common characteristic of star-forming regions.

We employ the $\oiii$/$\hb$ ratio to trace highly ionized regions and $\nii$/$\ha$, $\sii$/$\ha$, and $\oi$/$\ha$ ratios for regions of lower ionization. The higher $\oiii$/$\hb$ ratio is concentrated in star formation regions and decreases towards their peripheries. This aligns with the expectation that $\oiii$ is primarily generated by highly ionizing young stars. 

\begin{figure*}
    \centering
    \includegraphics[width=0.658\columnwidth]{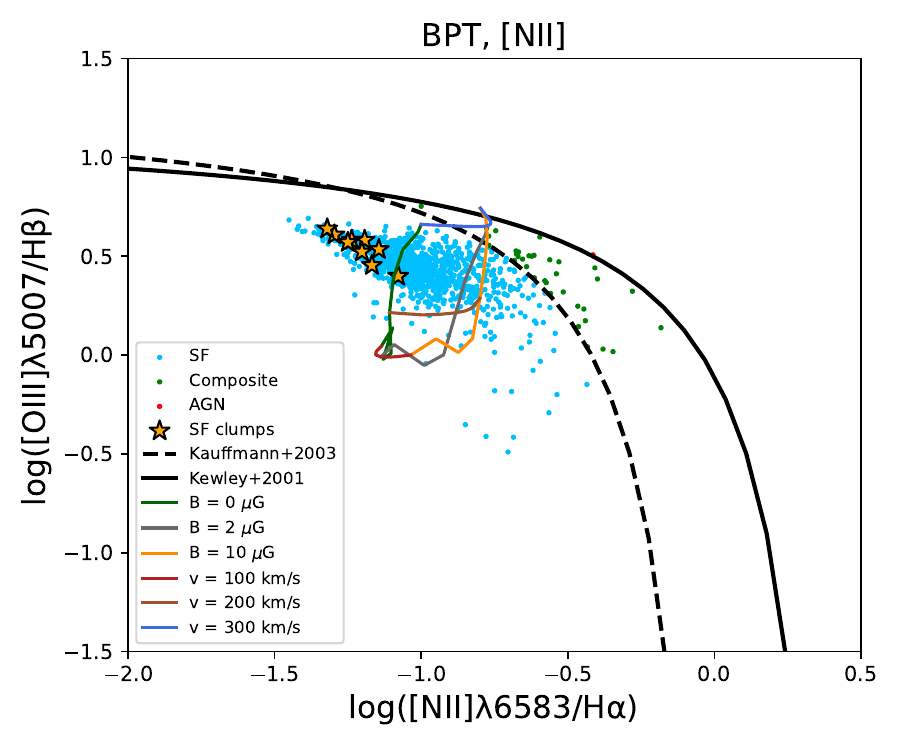}
    \includegraphics[width=0.658\columnwidth]{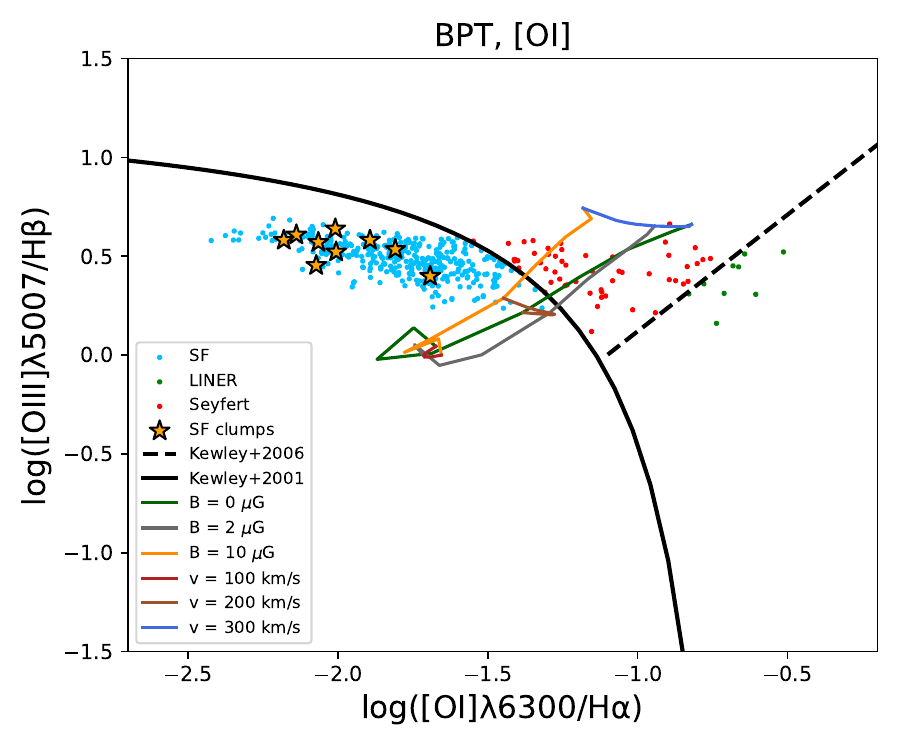}
    \includegraphics[width=0.658\columnwidth]{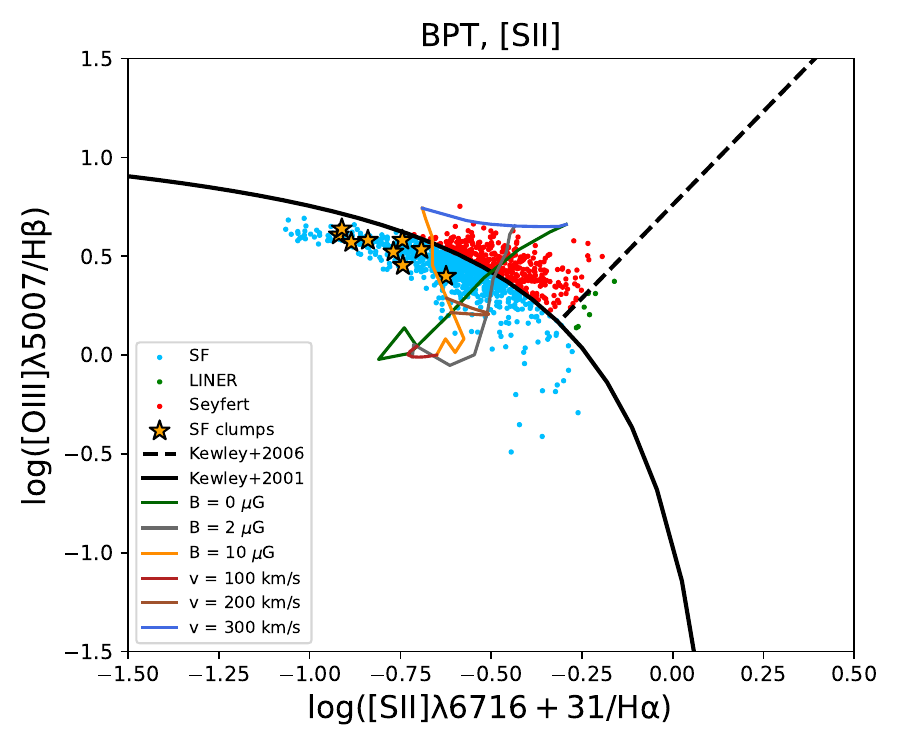}
    
	\includegraphics[width=1.96\columnwidth]{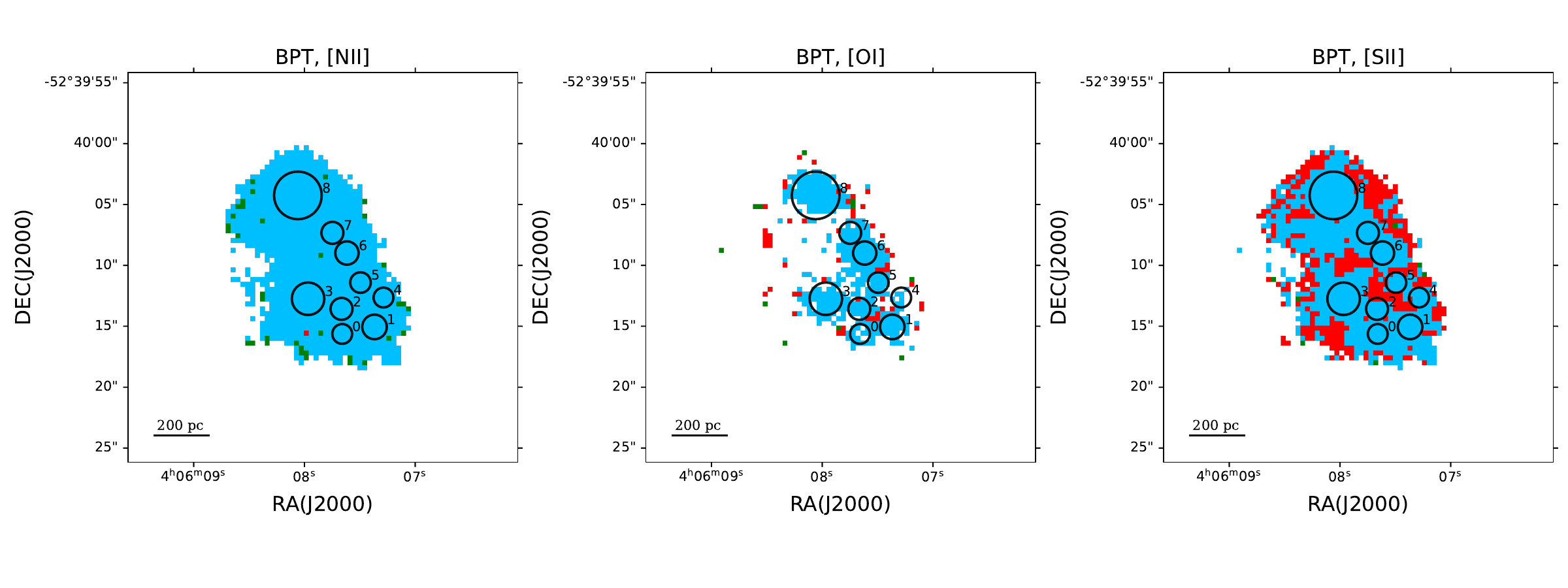}
    \caption{ The top panels display spatially resolved BPT diagrams using different emission-line ratios for NGC 1522. Solid lines, as defined by  \citet{2001ApJ...556..121K}, and dotted lines, as defined by \citet{2003MNRAS.346.1055K} and \citet{2006MNRAS.372..961K}, represent demarcation curves delineating regions associated with star formation, active galactic nuclei (AGNs), and Low Ionization Nuclear Emission Regions (LINERs). Overlaid are grids of the MAPPING III library of shock + precursor model \citep{2008ApJS..178...20A}. The firebrick, sienna and royalblue lines represent the shock models with speeds of 100, 200, and 300 \kms, respectively. The corresponding spatial distributions are shown in the bottom panels. Black circles in the bottom panel mark the locations of the identified star-forming clumps. The field of view of lower panels matches that of Figure~\ref{knots}.} 
    \label{BPT}
\end{figure*}

Although BPT-like diagrams \citep{1981PASP...93....5B,2003MNRAS.346.1055K,2001ApJ...556..121K,2006MNRAS.372..961K} were initially applied to central galactic regions or entire galaxies, we employ them here at the spaxel level to identify distinct ionizing process. We examine spatial distributions using BPT diagrams, including $\oiii\lambda 5007$/$\hb$ versus $\nii\lambda 6583$/$\ha$, $\oi\lambda6300$/$\ha$, and $\sii\lambda\lambda 6716,6731$/$\ha$, which are shown in Figure~\ref{BPT}. Taking into account the impact of shocks, we also present the results from the MAPPINGS III shock model library \citep{2008ApJS..178...20A} on our BPT diagram. The MAPPINGS III shock + precursor models are used, considering shock velocities ranging from 100 to 300 \kms, electron density of 1 $ \rm cm^{-3}$, and Large Magellanic Cloud (LMC)'s abundance. The magnetic parameter varies from 0 to 10 $\mu$G.
The $\nii$-based BPT diagram indicates that the ISM is predominantly ionized by ultraviolet (UV) photons from younger stars throughout the galaxy. However, we note that in the $\oi$-based and $\sii$-based BPT diagrams, some regions of photoionization from the active galactic nucleus AGN-like exist at the outer edges of star-forming clumps, which will be discussed in Section~\ref{chemical properties}. 
We can see that as the shock velocity increases, the shock model extends towards the AGN-like region, reaching our AGN-like region at about 250 \kms. The N2 index that we use to calculate metallicity and $\log(N/O)$ does not change much with the shock velocity. However, \oiii/\hb\ in O3N2 index is more susceptible to shock.

\subsection{Gas-phase metallicity}\label{Sec_metal}

The gas-phase metallicity provides insights into a galaxy's star formation history, gas inflow/outflow processes, and material exchange with the ISM \citep[e.g.,][]{Maiolino2019,Peroux2020}. Various methods exist for measuring metallicity, particularly oxygen abundance, in the optical band. Some methods utilize photoionization models to interpret the strength of strong emission lines, such as N2O2 and R23, to constrain metallicity. Direct methods, on the other hand, rely on measurements of electron temperature (T$_e$) and line ratios involving forbidden or recombination lines relative to the $\hb$ line. However, due to the limited spectral range of the MUSE instrument, we cannot observe $\oii\lambda 4363$ emission line, which is to measure the crucial electron temperature (T$_e$). Consequently, we derive metallicity using other strong emission line ratios. 

The O3N2 index \citep{1979A&A....78..200A} is defined as 
\begin{equation}
    \text{O3N2} \equiv \log \left(\frac{\oiii\lambda 5007/\hb}{\nii\lambda 6583/\ha} \right)
\end{equation}
Following the \citet{2013A&A...559A.114M} calibration of 
\begin{equation}
    12 + \log(\text{O/H}) = 8.505 - 0.221 \times \text{O3N2}
\end{equation}
for O3N2 ranges from -1.1 to 1.7, we derive the metallicity with a typical error of 0.08 dex.

The N2 index \citep{1994ApJ...429..572S,2000MNRAS.316..559R} is defined as
\begin{equation}
    \text{N2} \equiv \log \left(\frac{\nii\lambda 6583}{\ha} \right)
\end{equation}
Following the \citet{2013A&A...559A.114M} calibration based on the CALIFA data and literature data using $T_e$, the polynomial relationship of the calibrated metallicity is derived as
\begin{equation}
    12 + \log(\text{O/H}) = 8.667 + 0.455 \times \text{N2}
\end{equation}
with a typical error of 0.09 dex when N2 ranges from -1.6 to -0.2.

\begin{figure}
    \centering
    \includegraphics[width=0.9\columnwidth]{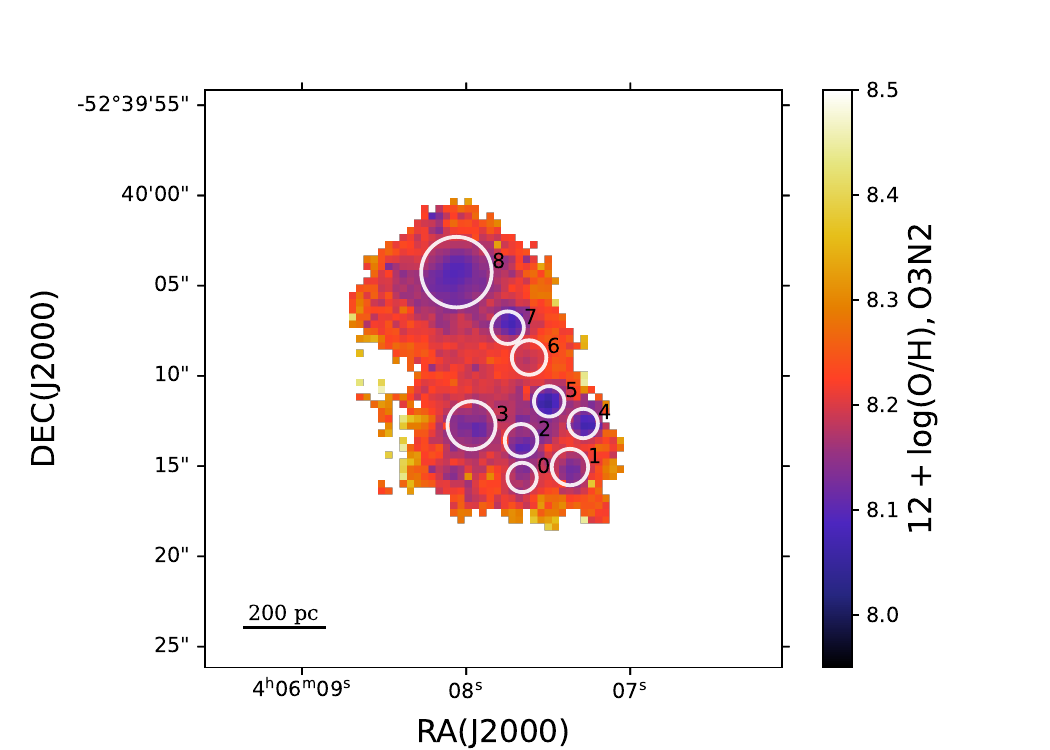}
    \includegraphics[width=0.9\columnwidth]{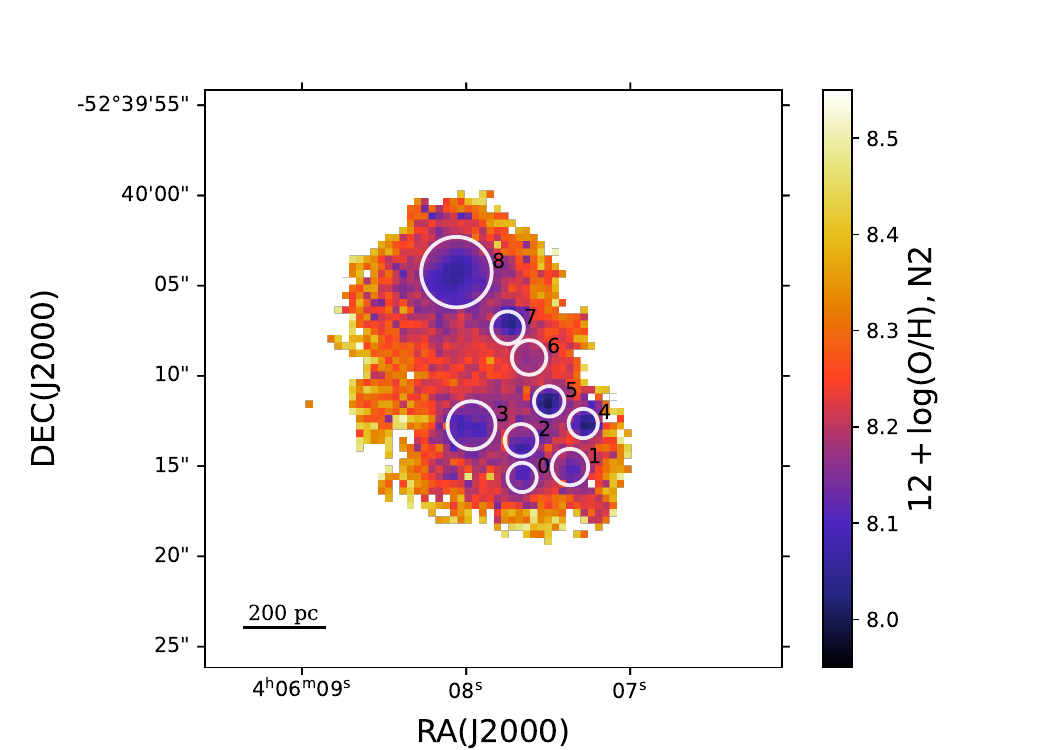}
    \caption{Metallicity map estimated from O3N2 (\textit{left panel}) and N2 (\textit{right panel}), respectively. The white circles indicate the locations of identified star-forming clumps. The field of view is the same as in Figure~\ref{knots}.}
    \label{metal}
\end{figure}

Figure~\ref{metal} displays the spatial distribution of gas-phase metallicity, revealing inhomogeneous metallicity patterns within NGC 1522. The most significant finding is that the metallicities of the nine star-forming clumps are all very low, approximately 0.2 to 0.3 dex lower than in the outer regions. The metallicity distribution varies significantly depending on the index calibrators. In the O3N2-based and N2-based maps, star-forming regions appear to exhibit lower metallicity, contrary to the prediction by \citet{2014MNRAS.437.3980P}, which suggested that the most active star-forming regions should have higher metallicity.

\subsection{N/O abundance}\label{subsec:figures}

In stellar nuclear processes, elements are categorized as either primary or secondary. In low-metallicity stars, carbon and oxygen serve as the building blocks for nitrogen production. Carbon and oxygen are synthesized through helium burning in galaxies, establishing a fixed ratio of nitrogen-to-oxygen (N/O). \citet{2021ApJ...908..183L} utilized the N/O abundance ratio to investigate the accretion of metal-poor gas in star-forming galaxies (SFGs). According to \citet{2019ApJ...872..144H}, when galaxies accrete metal-poor gas, an excess in N/O abundance should be observed, potentially triggering intense star formation processes \citep{2020ApJ...891...19L}.

However, the MUSE spectra do not cover the $\oii\lambda\lambda$3726,3729 lines. To calculate the N/O abundance ratio for each spaxel, we employ the empirical relation proposed by \citet{2009MNRAS.398..949P} instead of the N2O2 method used in \citet{2020ApJ...891...19L}. The relation is given by:
\begin{equation}
    \log (\text{N/O}) = 1.26 \times \text{N2S2} - 0.86.
\end{equation}
The uncertainty in $\log$(N/O) is determined from flux errors of the emission lines and systematic uncertainties in the N2S2 calibration. The N2S2 index \citep{2000ApJ...542..224D} is defined as
\begin{equation}
    \text{N2S2} \equiv \log \left(\frac{\nii\lambda 6583}{\sii\lambda\lambda 6716,6731} \right),
    \label{N2S2}
\end{equation}
which is sensitive to metallicity and weakly depends on the ionization parameter. 

\begin{figure*}
    \centering
    \includegraphics[width=0.9\columnwidth]{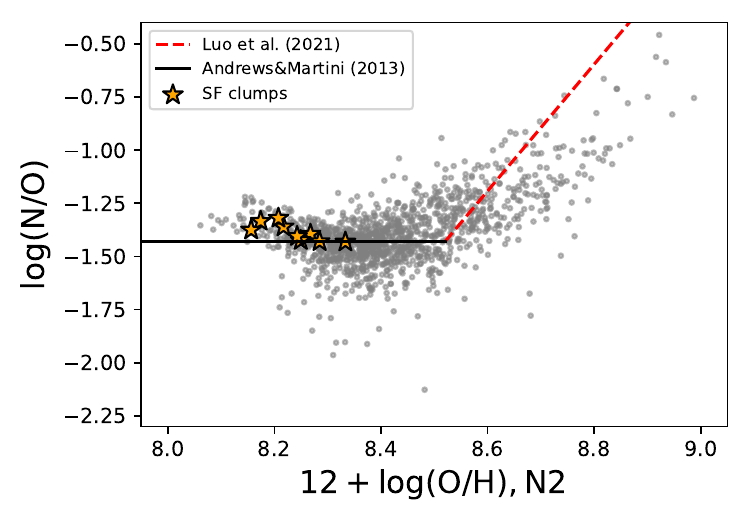}
    \includegraphics[width=1.05\columnwidth]{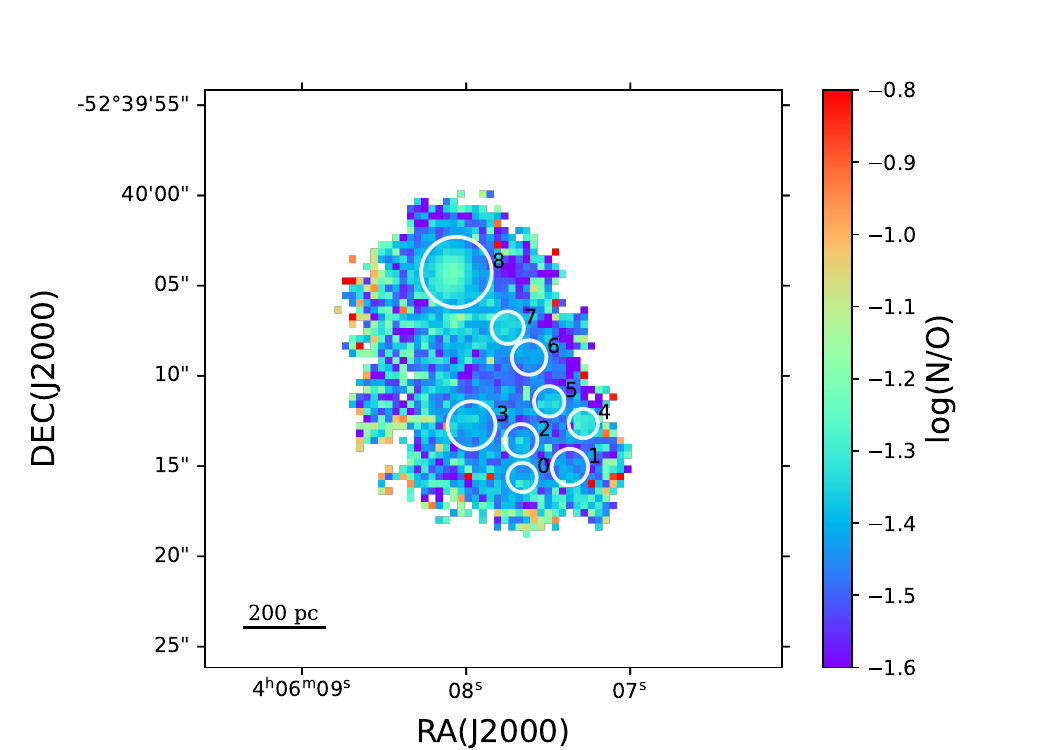}
    \caption{\textit{Left:} The N/O vs. O/H (N2) plane for NGC 1522. The grey dots are the spaxels from the galaxy. The orange star marks the star-forming clumps. The black solid line is the result from \citet{2013ApJ...765..140A}. The red dash line is the best-fit line for normal star-forming galaxies from \citet{2021ApJ...908..183L}. \textit{Right:} The spatial distributions of  $\log N/O$. The white circles indicate the locations of identified star-forming clumps. The field of view of the right panel is the same as in Figure~\ref{knots}. }
    \label{NO_OH}
\end{figure*}

The relationship between N/O and O/H  is plotted in the left panel of Figure~\ref{NO_OH}, where the black solid line is taken from \citet{2013ApJ...765..140A}. Additionally, we incorporate the fitting results from \citet{2021ApJ...908..183L}, indicating an increase in N/O with O/H. In the right panel, we plot the map of $\log$(N/O). Compared to the map of 12 + $\log$(O/H), the distribution of nitrogen-to-oxygen ratio  appears to be more flat.

\subsection{Star formation rate}\label{sec:SFR}

Star formation is a pivotal process in galaxy evolution, influencing stellar populations, chemical enrichment, and the surrounding medium's energy and pollution. The $\ha$ emission line, which traces ionized gas in the vicinity of young stars, is widely used to measure the SFR. To assess the relative strength of star formation activity in NGC 1522, we calculate the extinction-corrected SFR from the $\ha$ luminosity, $L(\ha)_{\text{cor}}$, using the calibration relation from \citet{1998ApJ...498..541K}:
\begin{equation}
    \text{SFR}(M_\odot \text{yr}^{-1}) = 7.92\ \times 10^{-42}\times L(\ha)_{\text{cor}}\ (\text{erg s}^{-1}) .
    \label{Eq.SFR}
\end{equation}
The uncertainty in the SFR contains uncertainties in $\ha$ luminosity, and attenuation. The total SFR of NGC 1522 is determined to be 0.098 $\pm$ 0.003 $\msun$ yr$^{-1}$. In a different approach, \citet{2014ApJ...783..135A} computed the total SFR of NGC 1522 by modeling the spectral energy distribution (SED) using UV/optical/NIR data from GALEX, 2MASS, Spitzer IRAC, and MIPS. Their modeling yielded an SFR value of 0.091 $\pm$ 0.023 $\msun$ yr$^{-1}$, which is consistent with our result.

\begin{figure}
    \centering
    \includegraphics[width=1\columnwidth]{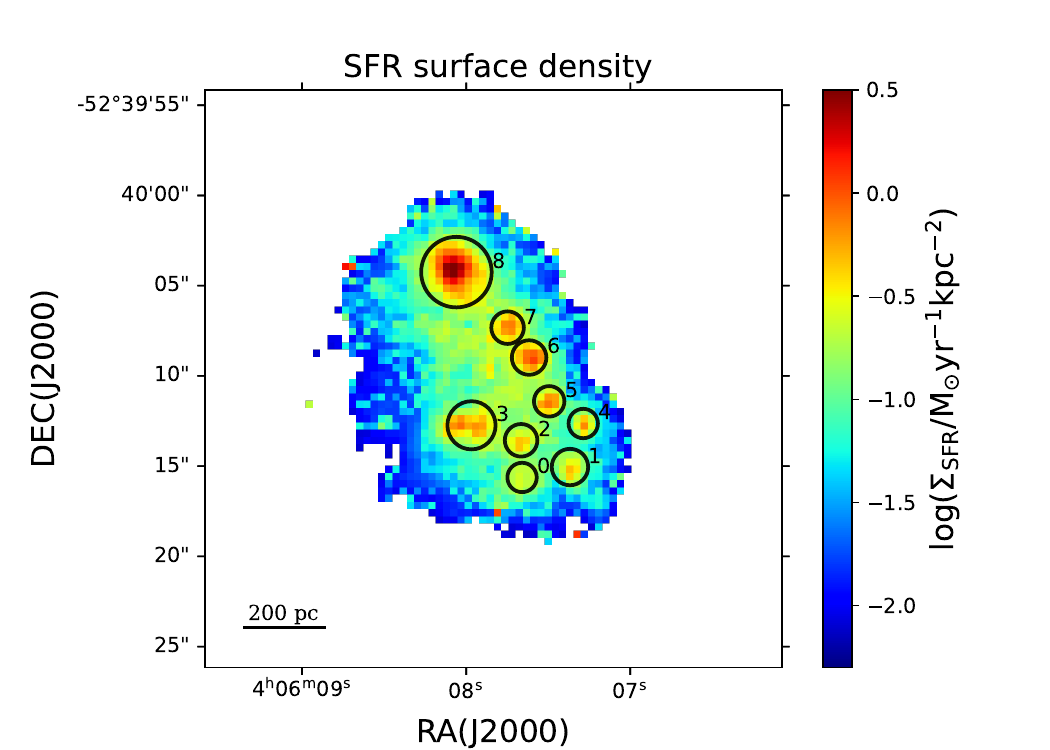}
    \caption{The spatial distribution of SFR surface density for NGC 1522 and the solid line presents the ellipse fitting of star-forming clumps. The black circles present the locations of identified star-forming clumps. The field of view is the same as in Figure~\ref{knots}.}
    \label{SFR}
\end{figure}

We also derive the SFR surface density ($\Sigma_{\text{SFR}}$) by dividing the total SFR from Eq.(\ref{Eq.SFR}) by the area of each region. The $\Sigma_{\text{SFR}}$ distribution in Figure~\ref{SFR} reveals that regions with the highest SFRs are situated in the northern part of NGC 1522. We find that these areas with enhanced SFR also show lower metallicity, as shown in Figure~\ref{metal}.

\subsection{Main Sequence Relation}
\begin{figure}
    \centering
    \includegraphics[width=\columnwidth]{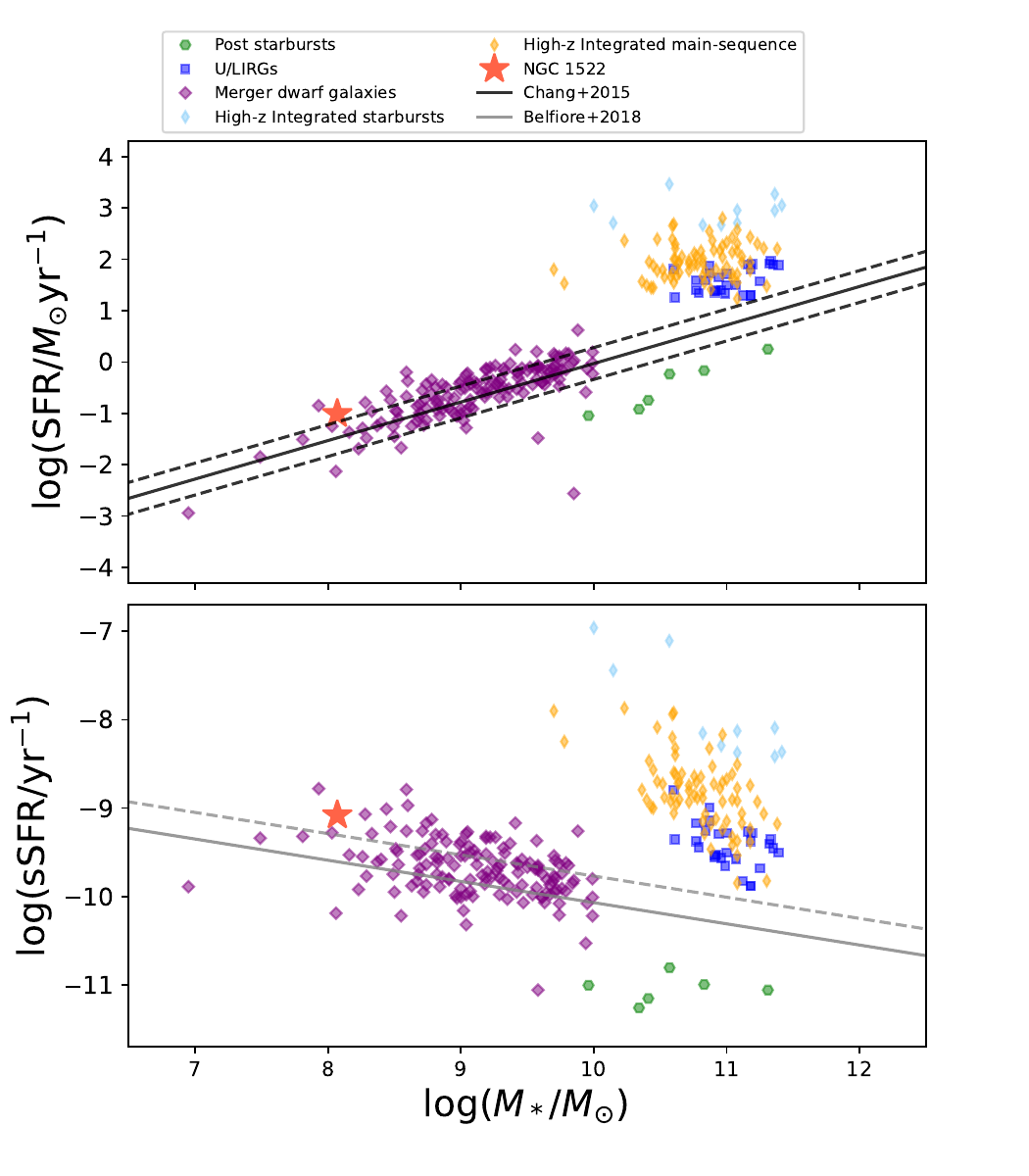}
    \caption{Top panel: SFR vs. stellar mass. The black solid line represents the main sequence relation from \citet{Chang_2015} where the dashed lines show the 1$\sigma$ scatter. Bottom panel: sSFR versus stellar mass. The gray solid line shows the best-fitting relation from \citet{2018MNRAS.477.3014B} where the dashed line is a factor 2 above this fit. The green hexagons are the post-starbursts. The blue squares are the massive merger (U)LIRGs. The purple diamonds represent the merging dwarf galaxies in the local universe. The sky-blue diamonds and orange diamonds are the high-z integrated starbursts and high-z integrated main-sequence galaxies, respectively. NGC 1522 is shown as a larger red star.}
    \label{fig:MSR}
\end{figure}

\begin{figure}
	\centering
	\includegraphics[width=\columnwidth]{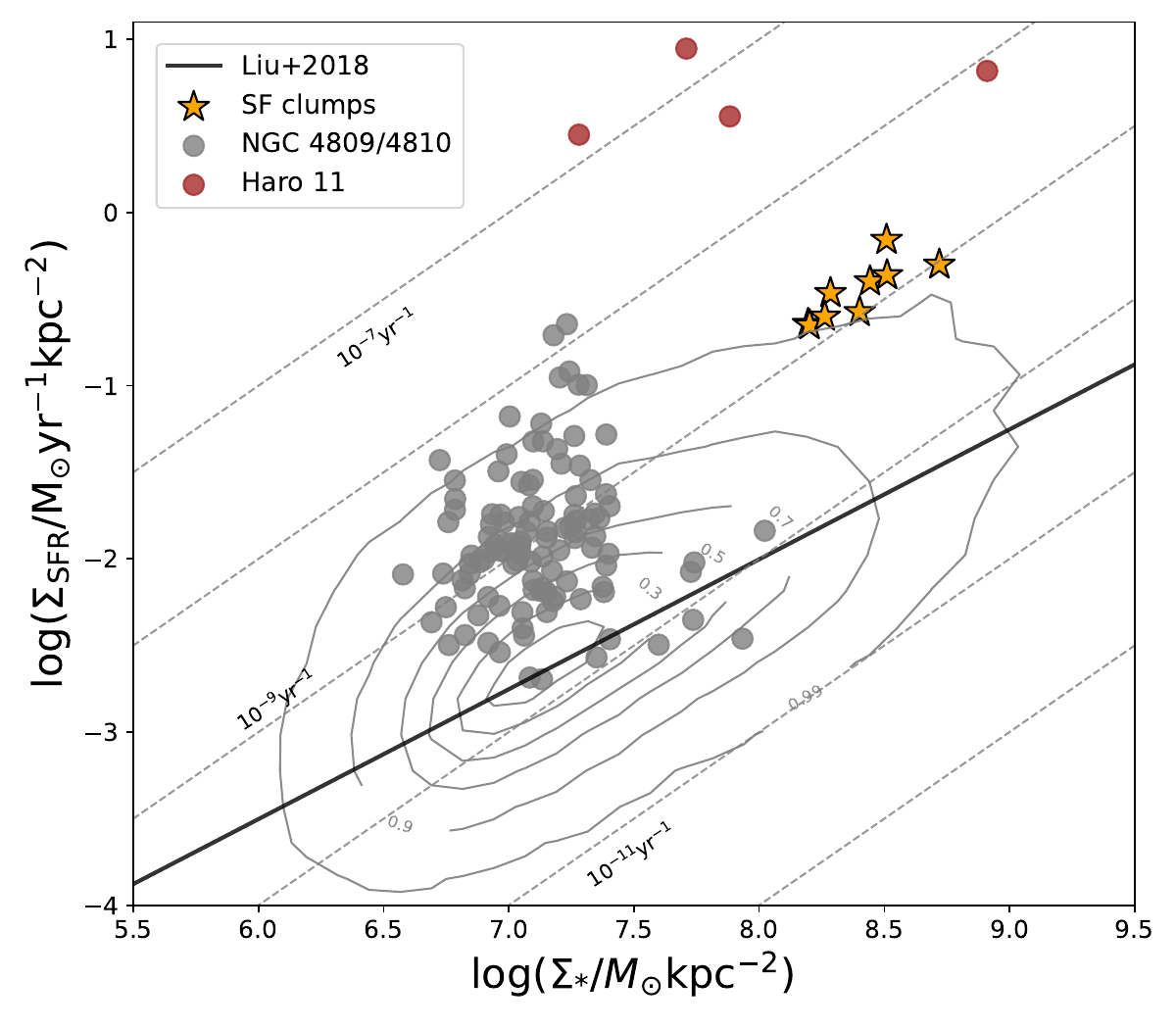}
	\caption{Stellar mass surface density vs. SFR surface density for SF clumps. The brown dots represent the SF clumps in Haro 11 from \citet{gao2022a}. The grey dots are the SF clumps in NGC 4809/4810 from \citet{gao2023}. The orange star marks the SF clumps in NGC 1522. The black solid line shows the spatially resolved main-sequence relation fited by \citet{2018ApJ...857...17L} where the dashed lines represent sSFR levels. The gray contours are the distribution of normal SF galaxies from the MaNGA survey.}
	\label{fig:rMSR}
\end{figure}

In this section, we compare the global star-forming activity of NGC 1522 with other galaxy samples. We derived the stellar mass based on the IRAC/Spitzer 3.6$\mu$m image following the literature method \citep{2008AJ....136.2782L}. To quantify the relative strength of star formation activities, we chose a comparison sample including ultraluminous
infrared galaxies((U)LIRGs) \citep{2019ApJ...870..104S}, merging dwarf galaxies in local universe \citep{2018ApJS..237...36P}, high-redshift starburst galaxies \citep{2018ApJ...853..149S}, high-redshift main-sequence galaxies \citep{2013ApJ...768...74T} and post-starburst galaxies \citep{2018ApJ...855...51S,2022ApJ...929..154S}. In Figure~\ref{fig:MSR}, we plot the SFR and specific SFR (sSFR) versus $M_*$ diagram for NGC 1522 and other galaxies. We find that NGC 1522 is higher than the main sequence relation and also higher than most of the merging dwarf galaxies. In the sSFR versus $M_*$ diagram, we consider a factor 2 above the best-fitting relation from \citet{2018MNRAS.477.3014B} as starbursts. NGC 1522 is located in the starburst region. NGC 1522 has similar sSFR to high-z galaxies. Besides, the spatially resolved main-sequence relation is shown in Figure~\ref{fig:rMSR}. We compare NGC1522 with two merger systems, NGC 4809/4810 and Haro 11 \citep{gao2022a,gao2023}. As an isolated dwarf galaxy, the star formation in NGC 1522 is more active than in NGC 4809/4810, which was identified as a merging system \citep{gao2023}. Obviously, these nine clumps can be considered as starbursts based on our criterion above. To account for the difference in spatial resolution between the IRAC/Spitzer image and the MUSE data, we convolved the MUSE SFR surface density map to match the resolution of the IRAC image. The convolution was performed using a Gaussian kernel with a full-width at half-maximum (FWHM) corresponding to the IRAC/Spitzer 3.6$\mu$m point spread function (PSF). We found minimal changes (less than 0.1 dex) in the derived SFR density distributions. This indicates that the differing spatial resolutions of the datasets do not significantly affect the trend presented in Figure~\ref{fig:rMSR}. The measurements shown in Figure~\ref{fig:rMSR} correspond to the values obtained before the PSF matching.


\subsection{Kinematics of ionized gas}

\begin{figure*}
    \centering
    \includegraphics[width=0.9\columnwidth]{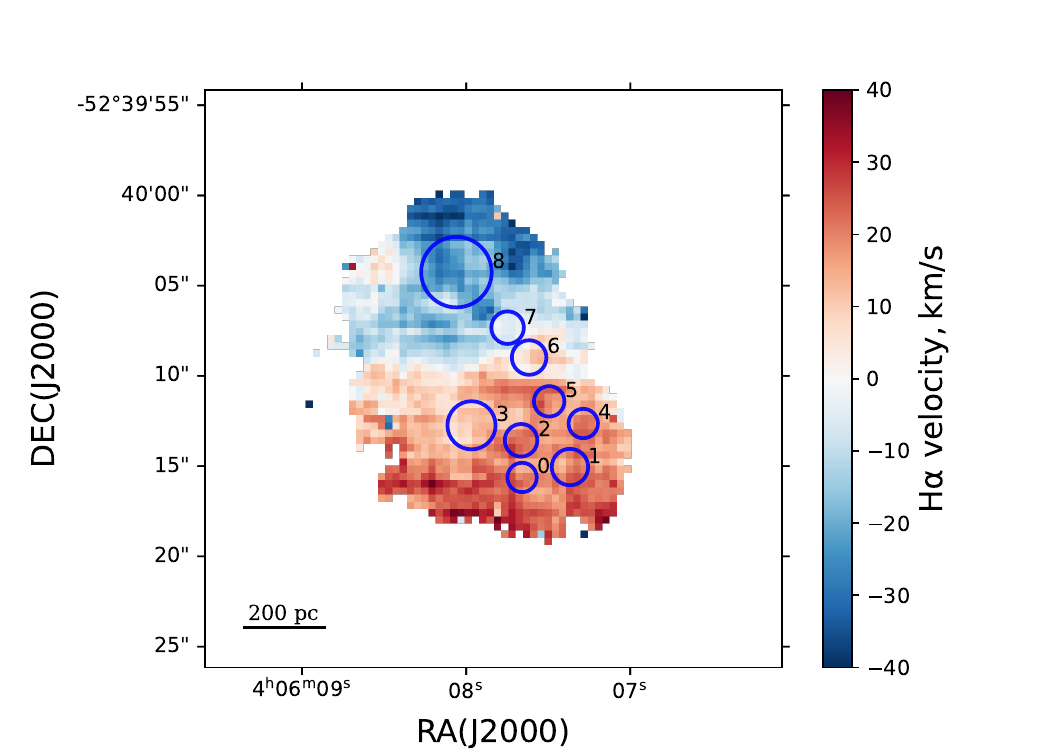}
    \includegraphics[width=0.9\columnwidth]{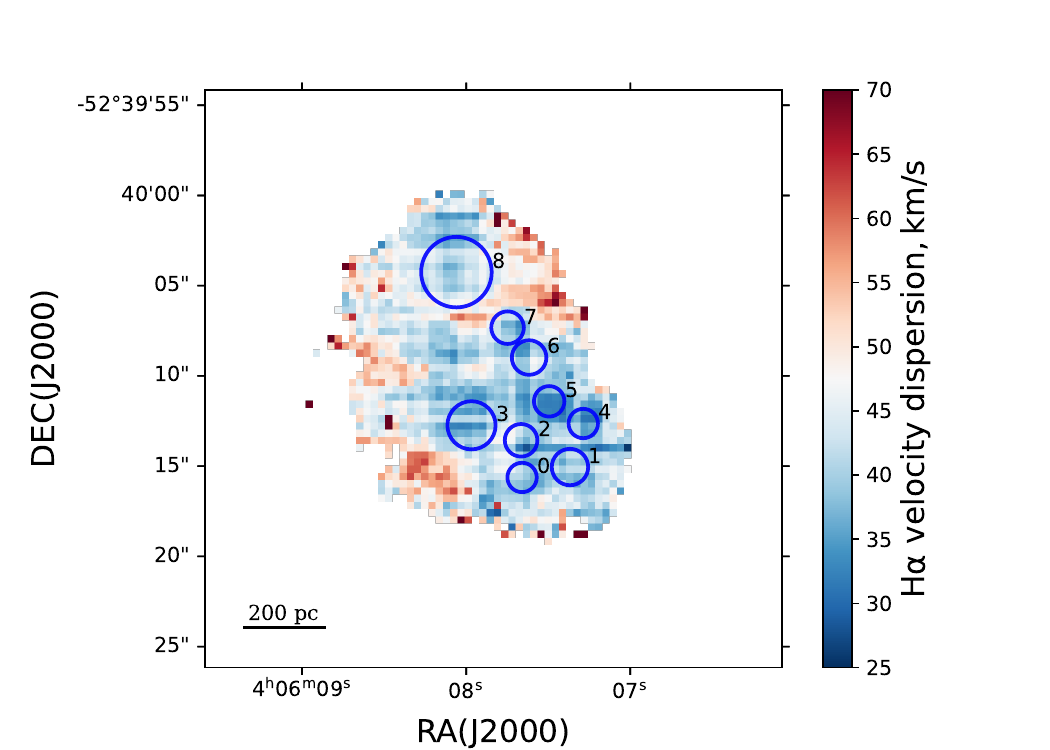}
    \includegraphics[width=0.9\columnwidth]{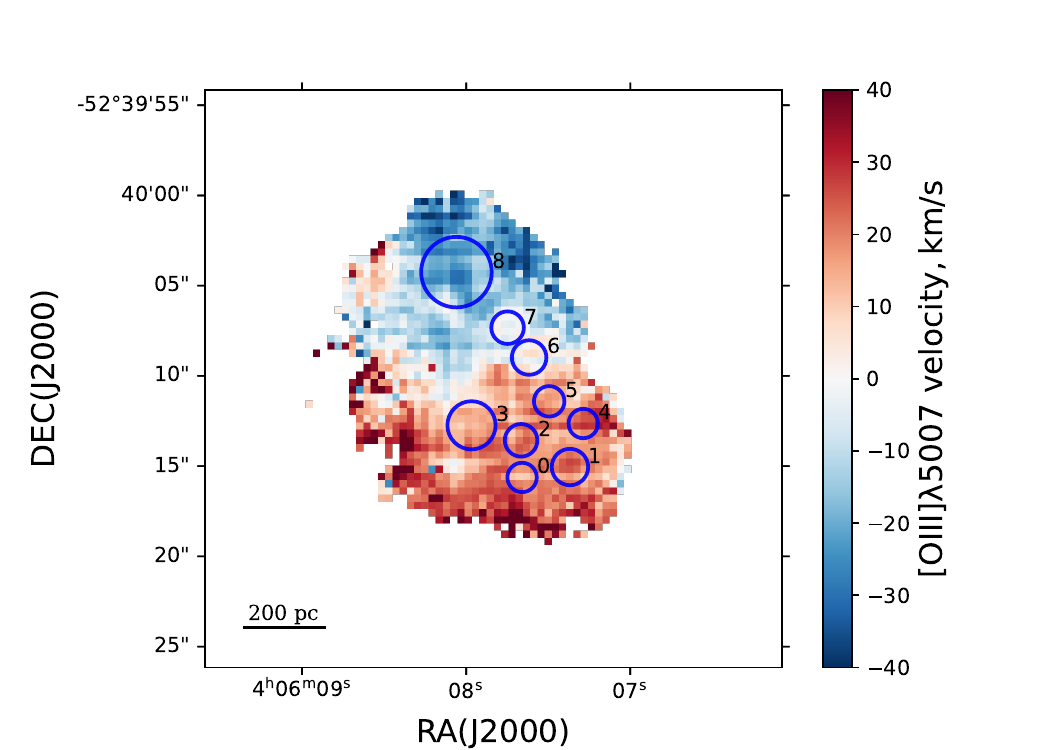}
    \includegraphics[width=0.9\columnwidth]{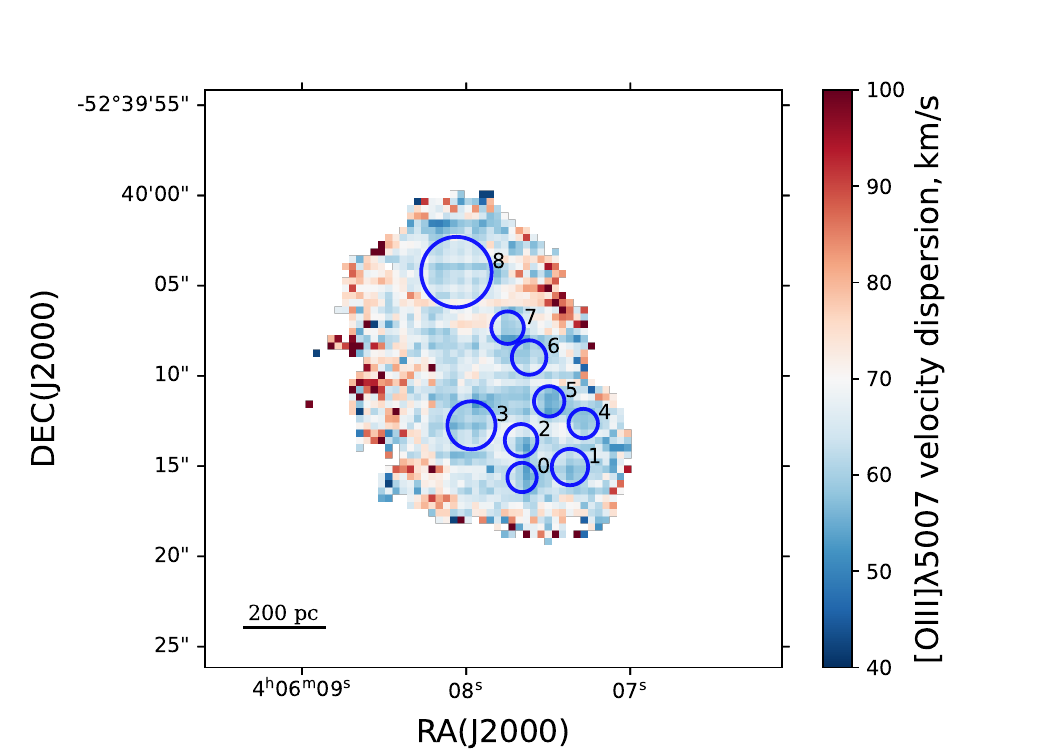}
    \caption{The spatial velocity field and velocity dispersion for $\ha$ (\textit{top panel}) and $\oiii\lambda$5007 (\textit{bottom panel}) kinematic components. Blue and red colors in the velocity maps correspond to blueshift and redshift. The blue circles present the locations of identified star-forming clumps. The field of view is the same as in Figure~\ref{knots}.}
    \label{velocity}
\end{figure*}

In Figure~\ref{velocity}, we depict the velocity maps of ionized gas traced by $\ha$ and $\oiii\lambda$5007 in the left panel. Contours in these maps are the same as in Figure~\ref{knots}. These velocity fields exhibit similar features, showing a clear increase in rotation velocity towards the outer region of the galaxy. The rotation axis aligns with the northeast-southwest direction, with maximum $\ha$ velocity around 40 km s$^{-1}$ and $\oiii\lambda$5007 velocity around 60 km s$^{-1}$.

Velocity dispersions of the ionized gas are calculated as $\sigma = \text{FWHM}/2.355$ of the $\ha$ and $\oiii\lambda$5007 emission lines, and we also corrected the instrumental broadening of MUSE which is 100 \kms. In the right panel of Figure~\ref{velocity}, we present the velocity dispersion distribution. $\ha$ velocity dispersion ranges from 17 to 130 \kms, while $\oiii\lambda$5007 velocity dispersion spans from 42 to 140 \kms. The highest velocity dispersion value locates on the west of knot (ID 8). The velocity dispersion map reveals regions with higher $\sigma$ values and \oi/\ha\ ratios, which do not align with the star-forming clumps. This might suggest that diffuse ionized gas (DIG) may be present in these regions. In Figure~\ref{velocity} we can see the presence of some stripes due to the data reduction. To minimize the impact, we applied a de-striping process to the data. While this process effectively removed the stripe artifacts, it also reduced the S/N. To assess the potential impact of this reduction on our results, we compared the velocity and velocity dispersion measurements in the nine clumps, finding that the de-striping process introduces a velocity difference of less than 10 \kms\ and a velocity dispersion  difference of less than 5 \kms. The galaxy maintains a regular rotational pattern, confirming that these stripes do not affect the overall conclusions of this study.

\section{Discussions}
\label{discussion}

In this section, we explore the star formation activities in NGC 1522, classified as a dwarf and peculiar S0 galaxy. Our study benefits from the high spatial resolution of MUSE, allowing us to examine the ongoing star formation. Furthermore, we can map the flux distribution of multiple emission lines across the galaxy and analyze emission-line flux ratios. NGC 1522 exhibits a high degree of metal poverty, implying that the externally accrete gas is driving the ongoing star formation.

\subsection{Chemical properties}
\label{chemical properties}

We investigate the chemical properties of NGC 1522, leveraging our emission line ratio maps to construct spatially resolved BPT diagrams. These diagrams provide insights into the ionization state and source of ionization within the galaxy. Figure~\ref{BPT} illustrates three BPT diagrams of NGC 1522, with key lines separating different ionizing mechanisms. 

We observe AGN-like line ratios in the $\oi$-based and $\sii$-based BPT diagrams, although the presence of AGNs is not reasonable in this context. While our study primarily focuses on star-forming clumps and assumes photoionization dominated conditions, the potential contributions from shocks and DIG cannot be neglected. Such deviant line ratios could be attributed to the low metallicities and the presence of shocks. 
Low metallicity may lead to a bias in the judgment from the BPT diagram. In low metallicity environments, ionized gas exposed to radiation from young stars is easier to form and maintain. \citet{2017MNRAS.466.3217Z}  find that, relative to a \hii\ region, \sii/\ha\ , \nii/\ha\ and \oi/\ha\ ratios are enhanced in DIG. Thus, the contribution from DIG emission could cause the \hii\ region to move towards the LI(N)ER region, as shown in our Figure~\ref{BPT}.
We use the N2 method to calculate the metallicity, so the metallicity observed in the DIG might be overestimated. The typical enhancement of \nii/\ha\ is about 0.2 dex \citep{2017MNRAS.466.3217Z}. The metallicity in our star-forming clumps is 0.2-0.3 dex lower than the outer regions. Following the \citet{2013A&A...559A.114M} calibration, an enhancement of 0.2 dex in N2 would not result in an increase in metallicity of more than 0.1 dex. Therefore, the enhancement of the line ratios in the DIG does not solely explain the metallicity difference found. As the same time, the shock process resulting from dynamic interactions or stellar winds may also help to explain this phenomenon. However, both our gas velocity and gas velocity dispersion are less than 100 \kms, which is too low to indicate shock. \citet{2008ApJS..178...20A} presented the most robust shock models, but the minimum shock velocity in their models is larger or equal to 100 \kms. We plot the results of these models on Figure~\ref{BPT}, where we find that the region influenced by low-velocity shocks does not overlap with the region corresponding to our AGN-like line ratios. Although in $\sii$-based BPT, when the shock velocity reaches 275 \kms, the shock model reach our AGN-like region. But we believe that this is beyond the range of low-velocity shock (usually less than 200 \kms).
Therefore, we consider that shocks do not play a dominant role in these regions. Despite the strong star formation observed in NGC 1522, no clear outflow signature has been identified. 
MUSE spectral resolution limits our ability to isolate different velocity components and constrain the gaseous kinematics. Determining the origin of these deviant line ratios is a fascinating avenue in the future study, and additional high quality data could help our understanding of the ionizing process.

In Section~\ref{Sec_metal}, we calculated gas-phase metallicity using various strong-line calibrators. Figure~\ref{metal} reveals that most star-forming clumps exhibit lower metallicity compared to the surrounding ISM. The O3N2-based and N2-based metallicity maps share similar distributions. All calibration maps indicate deficient metallicity in the $\hii$ regions, possibly indicating the accretion of external metal-poor gas. Differences in metallicity maps are discernible and may be attributed to sensitivity discrepancies among different metallicity calibrators \citep[e.g.,][]{2008ApJ...681.1183K,2016Ap&SS.361...61D}. 

By analyzing the relative emission of ions with varying ionization potentials, we can untangle the ionization structure of our star-forming clumps. The $\oiii/\oii$ line ratio, which is sensitive to metallicity, is typically used to derive the ionization parameter \citep{2002ApJS..142...35K,2004ApJ...617..240K}. While the MUSE spectra lack coverage of the $\oii\lambda$3727 emission doublet, we observe higher $\oiii$/$\hb$ ratios (Figure~\ref{emission map}) and higher SFRs (Figure~\ref{SFR}) in the star-forming regions than in the outer regions. These results might suggest a high ionization parameter \citep{2014Ap&SS.350..741D} in these regions. The gas-phase metallicity derived from the O3N2 index shows lower metallicity with a higher ionization parameter, further emphasizing the complexity of ionization processes. The N2 index, widely used for metallicity estimation \citep[e.g.,][]{2002MNRAS.330...69D,2005MNRAS.361.1063P}, exhibits similar spatial distributions to the O3N2 index. These different metallicity behaviors could arise from calibration differences among these indices.

\subsection{Accretion of metal-poor gas}

Star formation serves as a valuable tracer of metal production within galaxies, as metals are synthesized through stellar nucleosynthesis and ejected into the ISM by dying stars. Without efficient metal mixing, the metal density in the ISM should exhibit a direct proportionality to the SFR. Figure~\ref{metal} clearly illustrates chemical inhomogeneity within the ISM, with most star-forming regions exhibiting lower metallicity. Additionally, these star-forming clumps do not coincide with the photometric center of the galaxy.

\citet{2015MNRAS.449.2588P} highlighted the significant impact of shear and turbulence on metal distribution within galaxies. Shear processes transport momentum from large to small scales, while turbulence acts to mix metals. In the absence of other mechanisms, galaxies should exhibit inside-out negative metallicity gradient, with metallicity decreasing towards the outer regions. Gas flows, either within the galaxy or from external sources, play pivotal roles in chemical evolution. \citet{1980FCPh....5..287T} presents closed-box chemical evolution models, suggesting that previous star-forming episodes can contaminate the gas reservoir in a galaxy, resulting in each new star-formation episode having a higher metallicity than the previous one. Additionally, the proximity of most star-forming regions to the galaxy's center is attributed to gas inflow toward the deepest gravitational potential well, triggering star formation. Consequently, the lower metallicity observed in these star-forming clumps could be explained by distinct processes.

The lower ionization forbidden emission line $\oi\lambda6300$ can trace shock heating in the interstellar medium (ISM). In the $\oi$/$\ha$ line ratio map, regions with strong $\oi$/$\ha$ ratios predominantly surround the star-forming clumps, suggesting the possibility of shock heating. However, we believe that the shock velocity should exceed 100 \kms\ within regions of intense star formation. If we consider the potential impact of stellar winds, which can blow out the products of star formation in dwarf galaxies with shallow gravitational wells, this could also contribute to the observed metallicity distribution. But we only find anomalous metallicity distribution, while the N/O distribution appeared flat. Meanwhile we did not detect any broad-line component produced by shock or stellar wind in the optical spectra.

The abundances of elements offer constraints on galaxy evolution and chemical enrichment. The  N/O abundance excess suggests the possibility of accreting metal-poor gas \citep{2019ApJ...872..144H,2021ApJ...908..183L,2022MNRAS.509.1237X}. Lower metallicities have been observed in several starburst dwarf galaxies \citep[e.g.,][]{2013ApJ...767...74S,2014ApJ...783...45S}, which are relatively common at high redshift \citep[e.g.,][]{2012A&A...539A..93Q}. Figure~\ref{NO_OH} displays the relation between N/O and O/H abundances, positioning the star-forming clumps in regions of the N/O vs. O/H plane that imply the accretion of metal-poor gas. Visual inspection of the optical image reveals no signature of apparent merger or interaction surrounding NGC 1522. Therefore, we postulate that the inside-out positive metallicity gradient can be explained by external gas accretion. 

While strong winds generated by Wolf-Rayet stars can transport nitrogen-enriched gas and elevate N/O in the ISM \citep{2007ApJ...656..168L}, in our high-SNR spectra, we do not observe typical Wolf-Rayet emission features. Although we cannot completely rule out the presence of Wolf-Rayet stars, this suggests that the enhanced N/O ratios are likely due to other processes. Furthermore, the metal-rich gas mixing with relatively metal-poor gas at larger radii, as proposed by \citet{2015MNRAS.449..867B}, would increase O/H and result in star-forming regions having higher metallicity, which contradicts the results in Figure~\ref{metal}. Therefore, we posit that the accretion of metal-poor gas is the most likely explanation for the metal composition of star-forming clumps in NGC 1522.

\subsection{Spatially resolved star-forming activity}
\label{SFactivities}
\begin{figure}
	\centering
	\includegraphics[width=\columnwidth]{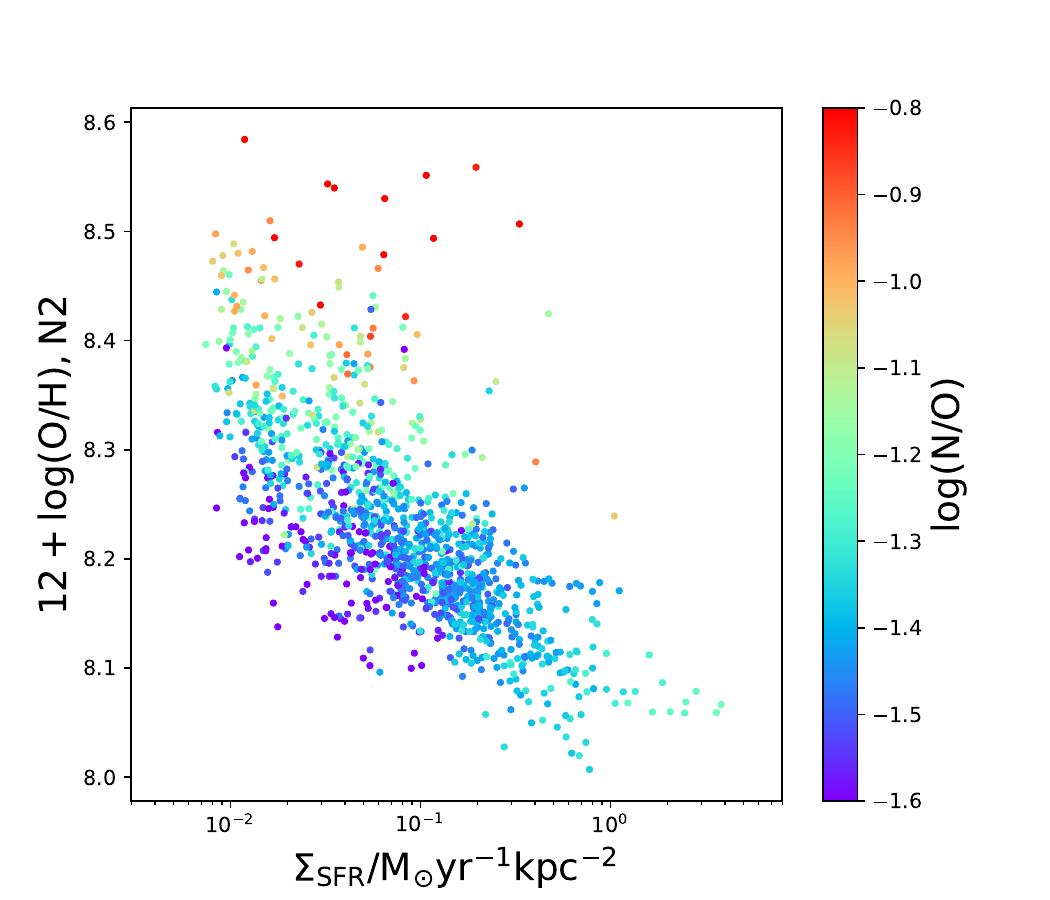}
	\caption{The metallicity(N2) vs. SFR surface density for each spaxel. The color presents the $\log$(N/O). The metallicity shows an anti-correlates with the SFR surface density, while the $\log$(N/O) shows no correlation with it.}
	\label{Z_sfr}
\end{figure}

In Section~\ref{SFknots}, we identified nine star-forming clumps in NGC 1522. We derived the physical properties of the radius in parsecs (pc), SFR and the gas-phase metallicity. To better illustrate the relationship between metallicity and star formation rate, we plot the metallicity(N2) vs. SFR surface densicy in Figure~\ref{Z_sfr}. We find that there is an anti-correlation between the metallicity and  SFR surface density, and no clear correlation between $\log$(N/O) and SFR surface density.
Although in Figure~\ref{Z_sfr}, we find some points with higher N/O ratios at both the very high and very low ends of SFR surface density. As shown in Figure~\ref{NO_OH}, these points are primarily concentrated in the largest and brightest star-forming clump, while the rest are distributed in the surrounding DIG regions. In the largest star-forming clump, intense star formation activity produces many young and massive stars. The nitrogen released during the evolution of massive stars can significantly increase the N/O ratio, which is known as self-enrichment. Unlike in \hii\ regions, the ionization parameters in DIG regions are typically much lower, resulting in a stronger presence of low-ionization stages (such as N$^+$ and S$^+$) \citep{1994ApJ...428..647D}. As a result, the N/O in these regions tends to be slightly higher.
\citet{2018MNRAS.476.4765S} discussed the possibility that external metal-poor gas fuels the star formation process, causing this anti-correlation between the metallicity and  SFR surface density. These phenomena can be explained by metal-poor gas accretion, which might be happening in NGC1522.

Notably, \citet{2012ApJ...752..111N} proposed that the overall structure and star formation history of high-redshift star-forming galaxies bear similarities to normal local star-forming galaxies, while the properties of individual star-forming regions resemble those of local nuclear starburst galaxies. NGC 1522 contains nine star-forming clumps which are all engaged in intense star-forming activity. These clumpy structure is also rare in the local universe. The majority of massive SFGs at redshifts around 1-3 also appear to be persistently fueled by gas, triggering ongoing star formation, rather than undergoing sporadic starbursts due to major mergers \citep{2011Msngr.145...39F}. Although the environment of NGC1522 is inevitably different from that of high-redshift galaxies, it is still an interesting local analogue to study.

\section{Conclusions}
\label{conclusion}

Our investigation of NGC 1522, an isolated dwarf galaxy, has yielded several significant findings. These findings shed light on the complex interplay of star formation, chemodynamics, and external influences in such galaxies. Here we summarise the main conclusions below.

\begin{enumerate}
    \item By leveraging IFU spectroscopic observations, we successfully identified nine distinct $\ha$ emission clumps within NGC 1522. These clumps represent regions where the dominant ionization source is star formation. It stands out as a galaxy with low metallicity, and a burgeoning population of young stars. 
    \item We meticulously extracted strong emission lines, including $\hb$, $\oiii$, $\oi$, $\ha$, $\nii$, and $\sii$, allowing us to construct spatially resolved line ratio maps. In addition, we have presented metallicity maps and BPT diagrams to investigate the underlying ionizing processes and the intriguing mixture of metal-poor and metal-rich ionized gas.
    \item The star-forming clumps exhibit lower metallicity levels and relatively flat nitrogen-to-oxygen ratios, implying an  inflow of metal-poor gas. This inflow of metal-poor gas into NGC 1522 appears to act as a trigger for, and sustainer of, ongoing star formation within this galaxy. We cannot rule out that some (possibly minor) contribution from DIG and self-enrichment could be present in regions where the N/O abundance ratio exceeds the average.
    \item Our analysis of velocity maps, specifically for the $\ha$ and $\oiii$ emission lines, unveiled a slow rotational pattern and minimal velocity dispersion. These observations collectively indicate the absence of merger events in NGC 1522's recent history.
    \item The star-forming clumps manifest a clumpy and off-center spatial distribution, reminiscent of high-redshift clumpy galaxies. This finding holds the potential to enhance our comprehension of remote star-forming entities.
\end{enumerate}

In summary, our study underscores the role of external gas accretion in driving starburst activities within isolated galaxies, exemplified by NGC 1522. These findings contribute to a deeper understanding of the intricate mechanisms shaping the properties of dwarf galaxies. To comprehensively unravel the molecular gas and dust components of NGC 1522, we emphasize the necessity of sub-millimeter observations, which hold the promise of unearthing further insights into the galaxy's evolution.

\begin{acknowledgments}
This work is supported by the National Natural Science Foundation of China (No. 12192222, 12192220 and 12121003). Y.L.G acknowledges the grant from the National Natural Science Foundation of China (No. 12103023). J.D. acknowledges the support of the National Science Foundation of China (NSFC) grant Nos. 12303010. This work is based on observations from the Multi Unit Spectroscopic Explorer (MUSE) at the Very Large Telescope (VLT), which is operated by the European Southern Observatory (ESO).
\end{acknowledgments}

\vspace{5mm}

\bibliography{new.ms}{}
\bibliographystyle{aasjournal}

\end{document}